\documentclass[aps,prd,amsmath,floats,floatfix, twocolumn,
superscriptaddress,nofootinbib,showpacs]{revtex4-1}

\usepackage[T1]{fontenc}
\usepackage[utf8]{inputenc}
\usepackage{lmodern}
\usepackage{tensor}
\usepackage{verbatim}
\usepackage{siunitx}
\usepackage[caption=false]{subfig}
\usepackage{wasysym}
\usepackage[english]{babel}
\usepackage{booktabs}
\usepackage{mathrsfs}
\usepackage{multirow}
\usepackage{makecell}
\usepackage[dvipsnames=true]{xcolor}
\usepackage[normalem]{ulem}

\definecolor{linkcolor}{rgb}{0.0,0.3,0.5}
\usepackage[
    hypertexnames=false,
    unicode,
    colorlinks=true,
    linkcolor=linkcolor,
    citecolor=linkcolor,
    filecolor=linkcolor,
    urlcolor=linkcolor,
    pdfusetitle
]{hyperref}
\usepackage{orcidlink}

\graphicspath{./Plots}
\begin{document}
    \title{Probing Intrinsic Ellipticity in Compact Star Binaries}
    \author{Zhiqiang Miao\orcidlink{0000-0003-1197-3329}}
    \affiliation{Tsung-Dao Lee Institute, Shanghai Jiao Tong University, Shanghai, 201210, China}
    \author{Huan Yang\orcidlink{0000-0002-9965-3030}}
    \email{hyangdoa@tsinghua.edu.cn} \affiliation{Department of Astronomy, Tsinghua University, Beijing 100084, China}
    \date{September 2025}
\begin{abstract}
We present a novel resonance mechanism that can occur in {compact-star} binaries: a spin-orbit resonance. 
This resonance locks the binary into a unique state where {the spin of one component} evolves alongside the orbit. 
The resonance requires this component to possess a finite ellipticity $\epsilon$, and we find that the locking probability is proportional to $\sqrt{\epsilon}$. 
We show that resonance locking and its subsequent breaking produce a characteristic phase signature in the gravitational waveform, opening a new observational channel for probing intrinsic ellipticity in compact-star binaries, including exotic compact objects. 
In addition, as an illustrative astrophysical scenario, we discuss magnetars, whose strong internal fields can source the required ellipticity and may place the signal in the ground-based low-frequency band, although their abundance at merger remains uncertain.
We have also conducted a search in all neutron star binaries up to the O4a gravitational-wave catalog, with no positive event found so far.
\end{abstract}
\maketitle

\section{Introduction}
Ten years after the first detection of binary black hole coalescence~\cite{LIGOScientific:2016aoc}, Gravitational Wave (GW) Astronomy has gradually become a precision science, with physical quantities measured at percentage levels, as particularly celebrated by the loudest event detected so far, GW250114~\cite{KAGRA:2025oiz}, at the signal-to-noise ratio (SNR) $\sim 80$. This not only allows for better and more frequent detection of events for population studies, but also the discoveries of small but important effects not necessarily included in current compact-binary templates, such as resonances~\cite{Lai:1993di,Schnittman:2004vq,Pan:2020tht,Poisson:2020eki,Kwon:2024zyg,Gerosa:2014kta,Miao:2025utd}, nonlinear ringdowns~\cite{London:2014cma,Cheung:2022rbm,Mitman:2022qdl,Cheung:2023vki,Khera:2023oyf,May:2024rrg,Khera:2024bjs,Ma:2024qcv}, memory~\cite{Zeldovich:1974gvh,Christodoulou:1991cr,Favata:2008yd,Favata:2009ii,Lasky:2016knh,Strominger:2014pwa,Yang:2018ceq}, etc.
{In this work, we point out an exciting opportunity to probe stellar ellipticity through a novel spin-orbit resonance mechanism, with particularly promising applications to neutron stars.}

Neutron stars are generically elliptical, as sourced from the internal magnetic stress. In fact, searching for continuous gravitational waves from rotating pulsars in our own galaxy is one of the science goals of ground-based detectors~\cite{LIGOScientific:2021mwx,LIGOScientific:2021ozr}. 
On the other hand, measuring star ellipticity $\epsilon$ has not been seriously considered previously, partially due to weak coupling of the ellipticity to the waveform, except in the scenario of exotic compact object (ECO) with order-unit deviation from spherical symmetry~\cite{Loutrel:2022ant}. 
We have discovered a novel spin-orbit resonance that naturally takes place in the inspiral stage, locking the evolution of spin and orbit for elliptical stars, providing a much stronger impact on the waveform. 
We also derive the resonance condition, the capture probability, the breaking frequency, and the associated waveform phase correction.

Although the mechanism itself is source agnostic, we discuss magnetars as one possible astrophysical realization, since strong internal magnetic fields can generate the ellipticity required for resonance locking.
We find that, under optimistic assumptions, third-generation (3G) detectors may marginally probe magnetar-like systems in the ground-based band, while decihertz detectors such as DECIGO~\cite{Kawamura:2011zz} may provide a more promising avenue.
In addition, a single detection likely leads to precise measurements of both the neutron star ellipticity and its moment of inertia, yielding significant implications for both astrophysics and fundamental physics. 
We have also performed parameter estimation of all binaries including neutron star up to the O4a GW catalog. The mass-gap event GW$190814$ shows an interesting peak (corresponding to $\epsilon \sim 10^{-2}$), but statistical evidence is weak. All other events present no signature of this spin-orbit resonance.

This paper is organized as follows. In Sec.~\ref{sec:mechanism}, we introduce the resonance mechanism and derive the resonance condition, capture probability, and breaking frequency. The corresponding gravitational-wave phase correction associated with resonance locking is derived in Sec.~\ref{sec:waveform}. Magnetars are discussed in Sec.~\ref{sec:magnetar} as an illustrative astrophysical realization of the mechanism. In Sec.~\ref{sec:LIGO-Virgo constraints}, we then search for resonance signatures in all neutron-star binaries reported up to the O4a gravitational-wave catalog. A population framework for interpreting future detections and non-detections is presented in Sec.~\ref{sec:population}. We conclude in Sec.~\ref{sec:discussion}.
Throughout this paper, we adopt the natural units $G=c=1$.

\begin{figure}%[h]--------------------------------------------------------------------------
\centering
\includegraphics[width=3.0in]{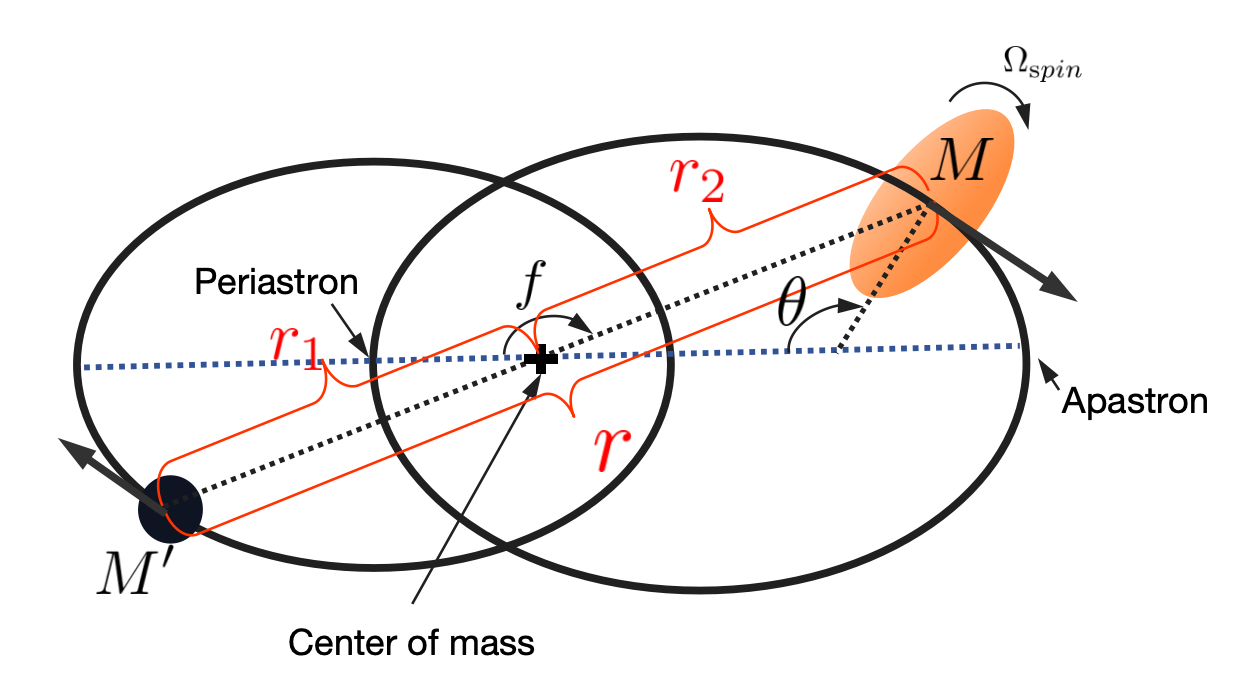}
\caption{Schematic diagram of the binary $M-M^\prime$ in the center-of-mass frame, where $M$ has a finite ellipticity. Here, $f$ is the angle between the line from the center of mass to $M$ and the line to the periastron, while $\theta$ is the angle between the line from the center of mass to $M$ and the long axis of the star. $r = r_1 + r_2$ represents the distance between $M$ and $M^\prime$.
}
\label{fig:illustration}
\end{figure}

\section{Resonance locking and breaking}
\label{sec:mechanism}
Consider a binary with an elliptical star, as illustrated in Fig.~\ref{fig:illustration}.
For simplicity, we merge the principal axis and the spin axis, and assume the spin is perpendicular to the orbital plane. 
We further restrict our discussion to circular orbits. 
(see Appendices~\ref{sec:ecc-orb} and~\ref{sec:misalign}  for discussions on general cases, including arbitrary spin–principal axis orientations and eccentric orbits.). 
The quadrupole moment of the star couples with the tidal field from the companion, which in turn affects the rotation of the star.
The equations of motion for the orbit and the star's rotation can be written as %\hy{define $I$ and other new variables}
\begin{align}
    &\ddot f +\frac32\epsilon\frac{M^\prime}{a^3}\frac{I}{I_{\rm orb}}\sin\left[2(f-\theta)\right] = {\dot\Omega_{\rm orb}}\,,\label{eq:fddot}\\
    &\ddot \theta -\frac32\epsilon\frac{M^\prime}{a^3}\sin\left[2(f-\theta)\right] = \dot \Omega_{\rm spin}\,,\label{eq:thetaddot}
\end{align}
where $M^\prime$ is the companion mass, $I$ and $I_{\rm orb}$ are the moments of inertia of the star and the orbit, respectively. $f$ is the true anomaly and $\theta$ is the rotation angle of the star, as shown in Fig.~\ref{fig:illustration}. 
For a circular orbit, $a = r$ represents the orbital separation.
$\dot\Omega_{\rm orb}$ is the orbital decay rate and $\dot\Omega_{\rm spin}$ is the neutron star spin down rate. We set $\dot \Omega_{\rm spin}$ to zero, since it is negligible compared to $\dot\Omega_{\rm orb}$, within the regime of our interest. We also neglect the $I/I_{\rm orb}$ term in Eq.~(\ref{eq:fddot}) since $I\ll I_{\rm orb}$. 
By defining the relative phase $\gamma = 2\theta-2f$, and combining Eqs.~(\ref{eq:fddot}) and (\ref{eq:thetaddot}) we obtain 
\begin{equation}\label{eq:pendulum}
    \ddot \gamma + \mathcal{B} \sin \gamma + \mathcal{C} = 0\,,
\end{equation}
where $\mathcal{B} = 3\epsilon M^\prime/a^3$ and $\mathcal{C}=  2\dot\Omega_{\rm orb}$.  

Equation~(\ref{eq:pendulum}) resembles the equation of motion of a rigid pendulum that is subject to a periodic torque and a (nearly) constant external torque. 
Such systems generally exhibit two types of motion: circulation and libration. When $\mathcal{B}>\mathcal{C}$, resonance locking becomes possible, driving the system into a libration state in which the stellar spin and orbit evolve synchronously.
The general theory of resonance capture predicts that the probability of entering resonance is approximately~\cite{Yoder:1979,Henrard:1982,Murray:2000} 
\begin{equation}\label{eq:probability}
    \mathcal{P}_{\rm lock} = \frac{2}{1+\frac{\pi}{2}\frac{\mathcal{B}^{1/2}\mathcal{C}}{\mathcal{\dot B}}} \approx \frac{4}{\pi} \frac{\mathcal{\dot B}}{\mathcal{B}^{1/2}\mathcal{C}}= \frac{4\sqrt{3}}{\pi}\left(\frac{q}{1+q}\right)^{1/2}\epsilon^{1/2}\,,
\end{equation}
where we have used the relation $\dot\Omega_{\rm orb}=-\frac{3}{2}\frac{\dot a}{a}\Omega_{\rm orb}$ and $q=M^\prime/M$ is the mass ratio. This is also confirmed by our numerical investigations.

On the other hand, when $\mathcal{B}<\mathcal{C}$, the resonance cannot be sustained and eventually breaks down. The corresponding gravitational-wave breaking frequency is
\begin{equation}\label{eq:fbr}
    f_{\rm br} = \frac{\Omega_{\rm br}}{\pi} = \frac{1}{\pi (M+M^\prime)}\left(\frac{5}{64}\frac{M+M^\prime}{M}\right)^{3/5}\epsilon^{3/5}\,.
\end{equation}

\begin{figure}%[h]--------------------------------------------------------------------------
\centering
\includegraphics[width=3.0in]{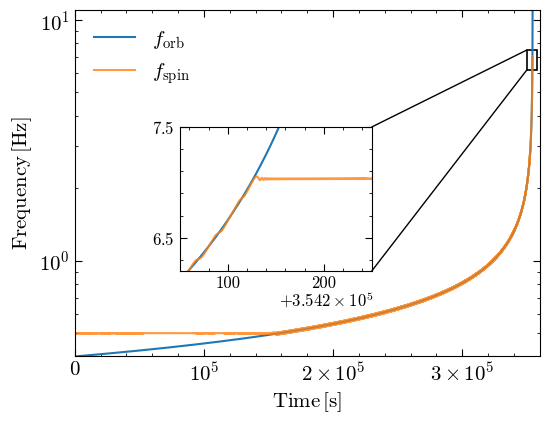}
\caption{The evolution of the orbital frequency and neutron star spin frequency with time for an illustrative NSBH system, where $M = 1.4\,M_\odot$, $q=3$, and $\epsilon = 5 \times 10^{-5}$. The initial phase is set as $f-\theta=0.3986$.}
\label{fig:evolution}
\end{figure}

In Fig.~\ref{fig:evolution}, we numerically evolve Eqs.~(\ref{eq:fddot}) and (\ref{eq:thetaddot}) for an illustrative neutron star-black hole (NSBH) system with an ellipticity in the range that may be produced by a strongly magnetized star.
We start with an orbital frequency $f_{\rm orb}=0.4\,{\rm Hz}$ and a stellar spin frequency $f_{\rm spin}=0.5\,{\rm Hz}$, and take $M=1.4\,M_\odot$, $q=3$ and $\epsilon=5\times10^{-5}$. 
As $f_{\rm orb}$ increases and passes through $0.5\,{\rm Hz}$, resonance occurs and $f_{\rm spin}$ begins to evolve synchronously with $f_{\rm orb}$. This locking continues until $f_{\rm orb}$ reaches about $7.1\,{\rm Hz}$ (corresponding to $f_{\rm br}=14.2\,{\rm Hz}$), where the locking breaks down.

\section{Waveform signature}
\label{sec:waveform}
Resonance locking induces angular-momentum exchange between the orbit and the elliptical star, which modifies the orbital decay and produces a phase shift in the gravitational waveform. 
Using the formula $d^2\psi/d\omega^2 = 2(dE/d\omega)/\dot E$ for the phase $\psi(f)$ of the frequency domain waveform $\tilde h(f) = \mathcal{A}e^{i\psi(f)}$~\cite{Tichy:1999pv}, we derive the phase correction due to resonance locking as
\begin{equation}\label{eq:phase corr}
    \delta\psi_{\rm lock}(f) = \left\{\!
\begin{aligned}
&-\frac{45}{64}\frac{I}{\mu^2M_t}(\frac{1}{v}-\frac{4}{3v_i}+\frac{v^3}{3v_i^4})\,, & f\leq f_{br}\,,\\ 
&-\frac{45}{64}\frac{I}{\mu^2M_t}(\frac{1}{v_{\rm br}}-\frac{4}{3v_i}+\frac{v_{\rm br}^3}{3v_i^4})\,, & f>f_{br}\,.
\end{aligned}
\right.
\end{equation}
where $M_t=M+M^\prime$ and $\mu = MM^\prime/M_t$ are the total mass and the reduced mass, respectively. 
Here $v= (\pi M_tf)^{1/3}$, $v_i<v_{\rm br}$ is an arbitrary constant related to the initial time and phase of the waveform. 
The correction clearly enters at 2PN (post-Newtonian) order, with a large prefactor $\sim I/(\mu^2M_t)\sim (R/M)^2$. 

It is useful to distinguish the locking-induced phase shift from the usual spin-related 2PN contributions, such as the spin–spin interactions and spin-induced quadrupole moments.
Although these effects enter at the same PN order, their amplitudes are controlled by the dimensionless spins, scaling schematically as $\chi_c\chi_\epsilon$ and $\chi_\epsilon^2$, respectively, where $\chi_\epsilon$ and $\chi_c$ denote the dimensionless spins of the elliptical star and its companion. 
For the low-spin systems relevant to resonance locking, $\chi_m \lesssim {\rm few}\times 10^{-2}$ as implied by the breaking frequency, these standard spin-related terms are therefore strongly suppressed.
Consequently, the resonance-locking effect is expected to dominate 
over these spin-related contributions in the parameter region of interest. 
More importantly, the resonance-locking effect introduces a characteristic frequency-dependent structure in the phase evolution that differs from those induced by spin-related effects [cf. Eq.~(\ref{eq:phase corr})].

\section{Magnetars as an illustrative astrophysical realization}
\label{sec:magnetar}
The resonance-locking mechanism derived above requires only a finite intrinsic ellipticity and can therefore be applied to a broad class of compact-object binaries. 
Magnetars provide one concrete astrophysical realization of the required ellipticity.
In the normal-conducting approximation, the ellipticity
can be estimated as~\cite{Gao:2022hzd}
\begin{equation}\label{eq:e-B}
\begin{split}
    \epsilon &= \kappa \frac{B_{\rm int}^2R^3}{M^2/R}\\ 
    &\approx 4\times10^{-6}\kappa\left(\frac{1.4\,M_\odot}{M}\right)^2\left(\frac{R}{12\,{\rm km}}\right)^4\left(\frac{B_{\rm int}}{10^{15}\,{\rm G}}\right)^2\,,
\end{split}
\end{equation}
where $M$ and $R$ are the mass and radius of the neutron star, respectively. $B_{\rm int}$ is the volume-averaged internal field.
Its value can be much larger than the surface magnetic field strength of magnetars (typically a few times $10^{14}\,{\rm G}$), and could even reach up to $10^{16}\,{\rm G}$, especially if a strong toroidal magnetic field is present~\cite{Ciolfi:2013dta,Mastrano:2011tf,Mastrano:2015rfa}.
The parameter $\kappa$ is of $\mathcal{O}(1)$ and depends on the internal magnetic field geometry and the equation of state (EOS) of the neutron star~\cite{Haskell:2007bh,Colaiuda:2007br,Lander:2009ib,Mastrano:2011tf,Mastrano:2015rfa}.

For this magnetic realization to be relevant during the late inspiral, the neutron star must retain a sufficiently strong internal field until merger. 
This requires both a short merger delay and sufficiently long internal-field retention~\footnote{Recently, a Fast Radio Burst (FRB) was reported to be located in a globular cluster~\cite{Kirsten:2021llv}. If this FRB is linked to a magnetar, it could imply that the activate lifetime of magnetars might be longer than previously thought~\cite{Kremer:2021zrj}.}.
The latter depends on the detailed magneto-thermal evolution of the star with uncertain microphysics~\cite{Goldreich:1992,Pons:2007vf,Aguilera:2007dy,Aguilera:2007xk,Glampedakis:2010sk,Lander:2012zz,Graber2015,Passamonti:2017pme} (see also Ref.~\cite{Pons:2025qzc} for a review), while the former is not expected for all isolated binary-evolution channels but may arise in dynamical scenarios such as hierarchical triples~\cite{Stegmann:2025clo,Liu:2018nrf}. 
Interestingly, recent analyses have reported evidence for nonzero eccentricity in the NSBH event GW200105, with $e_{20}\simeq 0.145$ at an orbital period of $0.1\,{\rm s}$~\cite{Morras:2025xfu,Kacanja:2025kpr,Jan:2025fps}, suggesting a possible dynamical origin. 
In such short-delay systems, a newly born magnetar could in principle retain a large enough internal field to source the ellipticity required for resonance locking.
A complementary motivation comes from short-GRB precursor flares, which are often interpreted as evidence for strong pre-merger magnetic interactions involving at least one highly magnetized neutron star~\cite{Suvorov:2024cff}. The occurrence rate of such precursors, at the level of a few percent, suggests that a small fraction of compact-binary mergers may involve magnetar-like components.

The resonance dynamics in this magnetar scenario is particularly simple to interpret.
Magnetars are typically observed with spin periods of order $1$--$10\,{\rm s}$, corresponding to spin frequencies $f_{\rm spin}\sim 0.1$--$1\,{\rm Hz}$.
The resonance therefore begins when the orbital frequency sweeps through this range. Once captured into locking, the system can remain locked until the breaking frequency given by Eq.~(\ref{eq:fbr}),
\begin{equation}\label{eq:fbr_mag}
    f_{\rm br} \approx 10\left(\frac{1}{1+q}\right)^{2/5}\left(\frac{1.4\,M_\odot}{M}\right)\left(\frac{\epsilon}{10^{-5}}\right)^{3/5}\,{\rm Hz}\,.
\end{equation}
Thus, if the neutron star retains a sufficiently strong internal field, $B_{\rm int}\gtrsim 10^{15}\,{\rm G}$, the induced ellipticity can be large enough for the locking phase to extend into the low-frequency band of ground-based detectors.
For smaller ellipticities, the resonance breaks at lower frequencies and may instead be more naturally probed by decihertz detectors such as DECIGO~\cite{Kawamura:2011zz}.

In what follows, we use ground-based detectors as a benchmark case and consider representative examples in which $f_{\rm br}$ lies inside their sensitive band.
A detailed investigation of the DECIGO regime is left for future work.
To assess whether such a signal can be reconstructed if present, we perform injection studies using the design sensitivities of the upcoming Advanced LIGO upgrade (A+)~\cite{LIGOAplus} and 3G detector network~\cite{KAGRA:2013rdx} (Cosmic Explorer~\cite{LIGOScientific:2016wof,Reitze:2019iox,Evans:2021gyd} and Einstein Telescope~\cite{Punturo:2010zza,Hild:2010id,ETdesign}; CE+ET).
We consider two representative systems: a NSBH system with $M^\prime=5M=7\,M_\odot$ and an equal-mass BNS system with $M^\prime=M=1.4\,M_\odot$, both placed at a luminosity distance $d_L=100\,{\rm Mpc}$ and the sky location of GW170817~\cite{LIGOScientific:2017ync}.
We adopt APR model~\cite{Akmal:1998cf} as fiducial EOS, which corresponds to a tidal deformability of $\Lambda=248$ and a moment of inertia of $I=30.18\,M_\odot^3$ for a $1.4\,M_\odot$ neutron star.
Other injection details are provided in Appendix~\ref{sec:Injection}.

\begin{figure*}%[h]--------------------------------------------------------------------------
\centering
\includegraphics[width=1.67in]{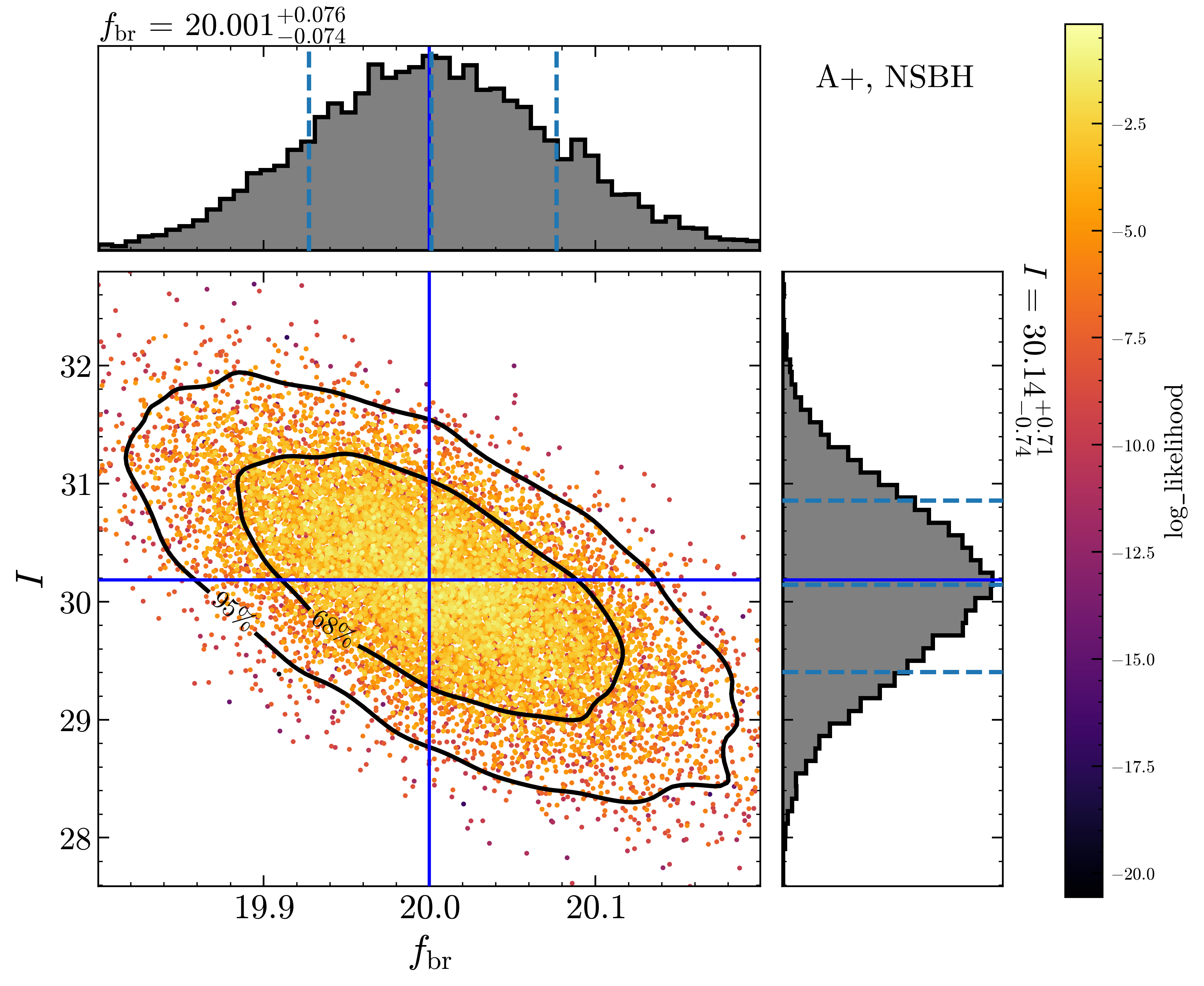}
\includegraphics[width=1.75in]{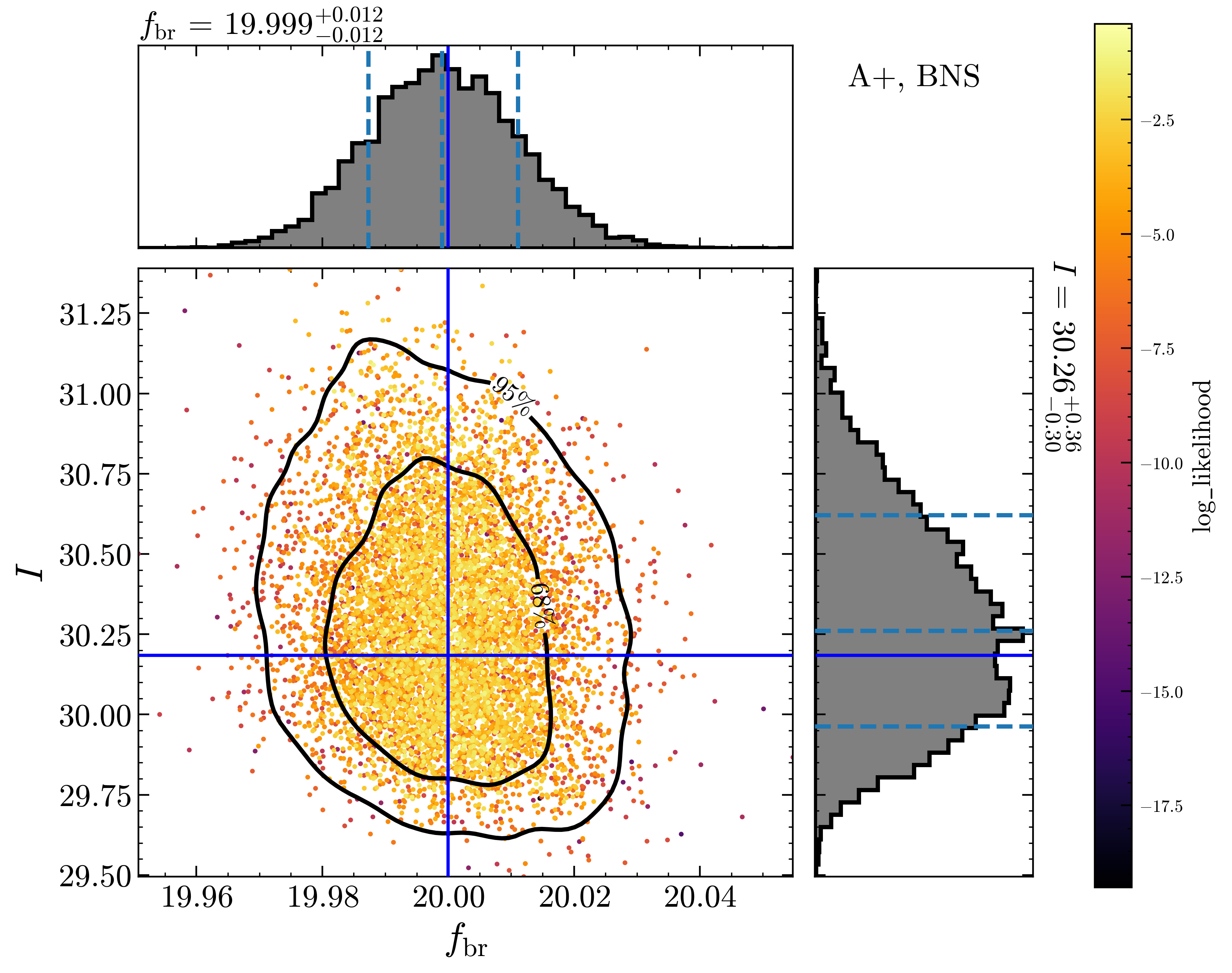}
\includegraphics[width=1.75in]{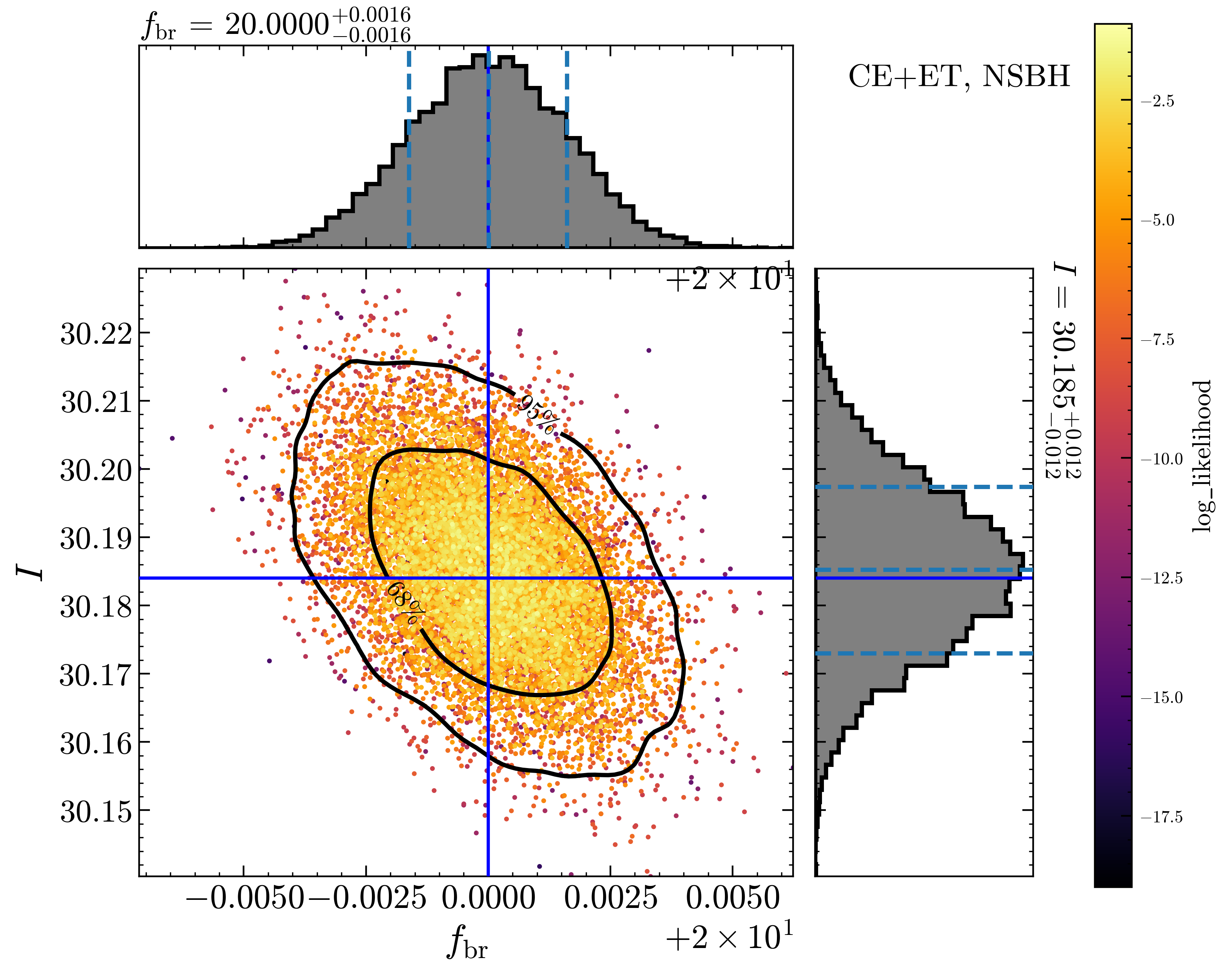}
\includegraphics[width=1.75in]{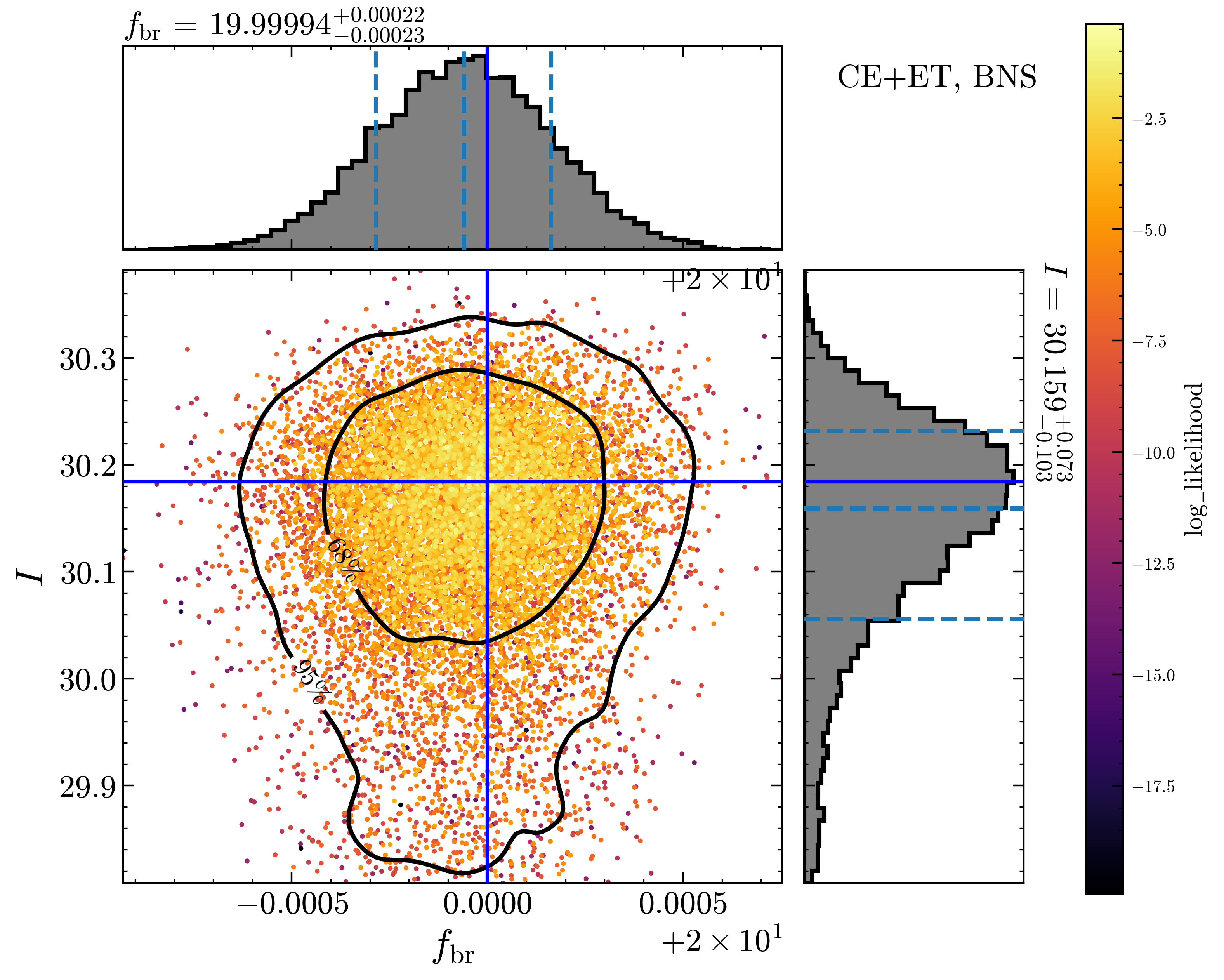}
\caption{Posterior distributions of breaking frequency $f_{\rm br}$ and moment of inertia $I$ for different recovered injections. The injection values are marked with blue lines.
}
\label{fig:Inject}
\end{figure*}

We first inject a locking signal with $f_{\rm br}=20\,{\rm Hz}$.
Figure~\ref{fig:Inject} shows the posteriors of $f_{\rm br}$ and $I$, demonstrating that all injected values are well reconstructed. 
For A+, the measurement accuracies of $f_{\rm br}$ and $I$ can reach $0.1\%$ and $1\%$, respectively; for CE+ET, they can be further improved by up to an order of magnitude.
We also examine a lower breaking frequency of $f_{\rm br}=10\,{\rm Hz}$ (see Appendix~\ref{sec:Injection}). 
In this case, CE+ET still recovers the injected parameters, while A+ does not. 
The reason is that the SNR accumulated in the $5-10\,{\rm Hz}$ band is only $\mathcal{O}(1)$ for A+, providing essentially no information, whereas CE+ET collects a SNR of $\mathcal{O}(100)$ in the same frequency band, enabling precise measurements.

% \section{Searches}

\begin{figure}%[h]--------------------------------------------------------------------------
\centering
\includegraphics[width=3.0in]{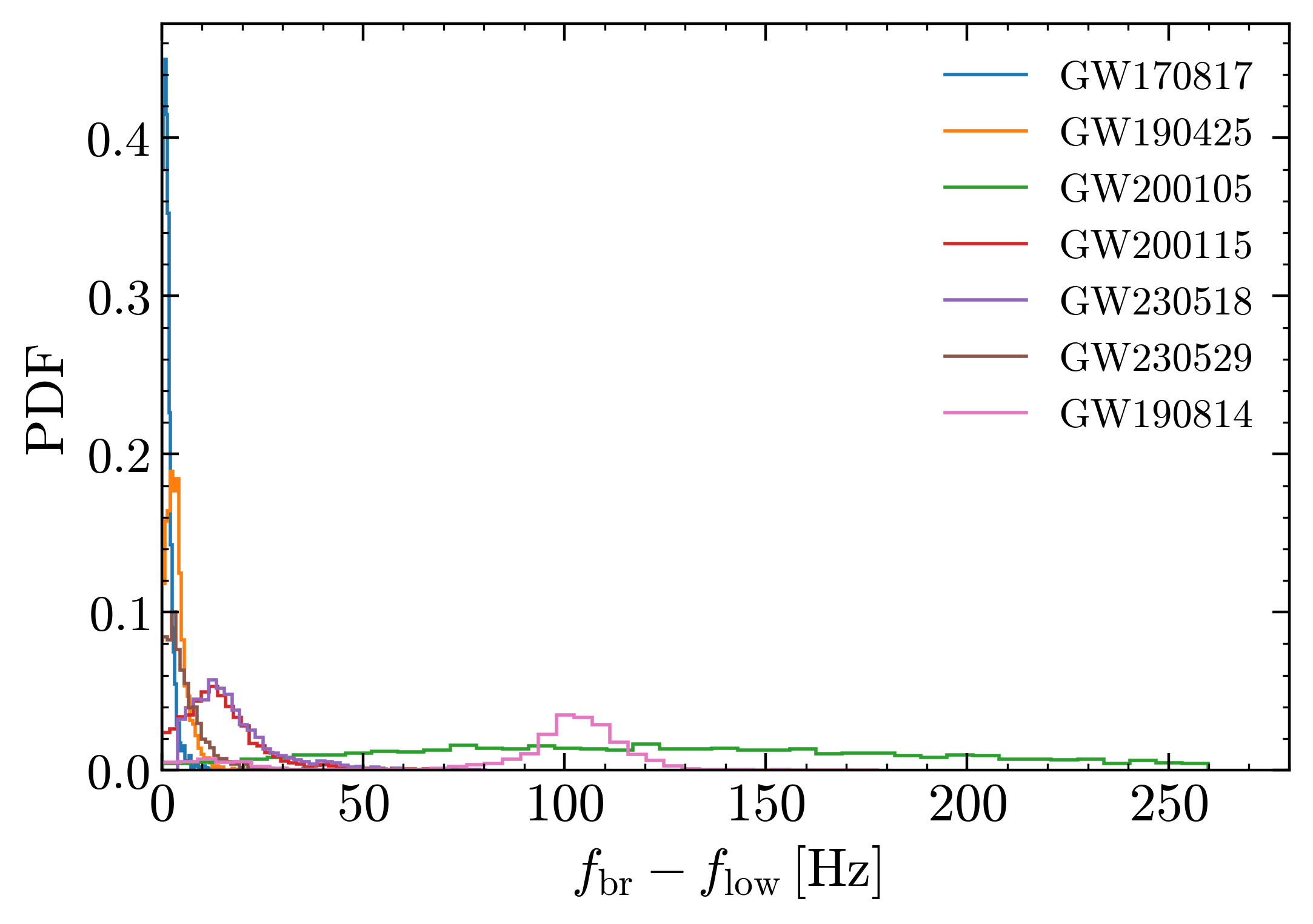}
\caption{Posterior distributions of breaking frequency $f_{\rm br}$ for the search in several GW events. Here, $f_{\rm low}$ indicates the minimum frequency of our search range. 
The distribution for GW200105 has been rescaled for visualization purposes.
}
\label{fig:search}
\end{figure}

\section{Constraints from LIGO-Virgo data}
\label{sec:LIGO-Virgo constraints}
We perform a first search for resonance-locking signatures in existing gravitational-wave events containing neutron stars. These include four NSBH systems (GW200105 and GW200115~\cite{LIGOScientific:2021qlt}, GW230518~\cite{LIGOScientific:2025slb}, GW230529~\cite{LIGOScientific:2024elc}), two BNS systems (GW170817~\cite{LIGOScientific:2017vwq} and GW190425~\cite{LIGOScientific:2020aai}) and one system with a $2.6\,M_\odot$ mass-gap object (GW190814~\cite{LIGOScientific:2020zkf}). 
For this analysis, we have used publicly available GW strain data~\cite{LIGOScientific:2025snk} and noise PSDs and followed an inference procedure similar to that conducted by LIGO~\cite{LIGOScientific:2014pky}, Virgo~\cite{VIRGO:2014yos}, and KAGRA~\cite{KAGRA:2020tym} (LVK) collaboration.

Figure~\ref{fig:search} presents the posteriors of $f_{\rm br}-f_{\rm low}$ for the seven systems. 
Here, $f_{\rm low}$ indicates the minimum frequency of our search range, which is $23\,{\rm Hz}$ for GW170817 and $20\,{\rm Hz}$ for the other systems.
It is obvious that no significant sign for $f_{\rm br} > f_{\rm low}$ is observed in both NSBH and BNS systems. 
In the hypotheses test framework, if we denote the absence of a locking signal as model $\mathcal{H}_0$ and the presence of a locking signal as model $\mathcal{H}_1$.
{We then compute the Bayes factor, $\mathcal{B}_0^1$, between the two models.}
For the four NSBH and two BNS systems, we obtain $\ln\mathcal{B}^1_0 \in [-2.99,-0.59]$, indicating that there is no preference for model $\mathcal{H}_1$ relative to model $\mathcal{H}_0$~\cite{Kass:1995}.

In the case of GW190814, the combined Hanford-Livingston-Virgo (HLV) analysis yields a bimodal posterior for $f_{\rm br}>f_{\rm low}$, as shown in Fig.~\ref{fig:search}. The left peak is consistent with the absence of a locking signal, while the right peak corresponds to a feature near $f_{\rm br}\sim 120\,{\rm Hz}$. 
However, the Bayes factor for the HLV data does not provide significant support for the locking model, with $\ln\mathcal{B}_0^1=-0.41$. We therefore regard this feature as statistically inconclusive. Additional single-detector checks are presented in Appendix~\ref{sec:searches}.
{If the $f_{\rm br} \sim 120\,{\rm Hz}$ feature were confirmed, it would be consistent with the secondary object of GW190814 being a neutron star rather than a black hole. 
In that case, the required ellipticity would be $\epsilon \sim 8\times10^{-3}$, corresponding to an internal magnetic field of 
${B}_{\rm int}\sim 8\times10^{16}\,{\rm G}$ [taking $\kappa=1$ in Eq.~(\ref{eq:e-B})].
Such a field strength is extreme for a neutron star and would place strong demands on the survival of the internal magnetic field until merger.}
Nevertheless, it remains below the virial limit 
($\sim10^{18}\,\rm{G}$) for neutron stars~\cite{Lai1991}. 
While the statistical evidence (the Bayes factor) is inconclusive, we conclude that this explanation is physically possible but not likely.

\section{Population studies}
\label{sec:population}
\begin{figure}%[h]--------------------------------------------------------------------------
\centering
\includegraphics[width=3.0in]{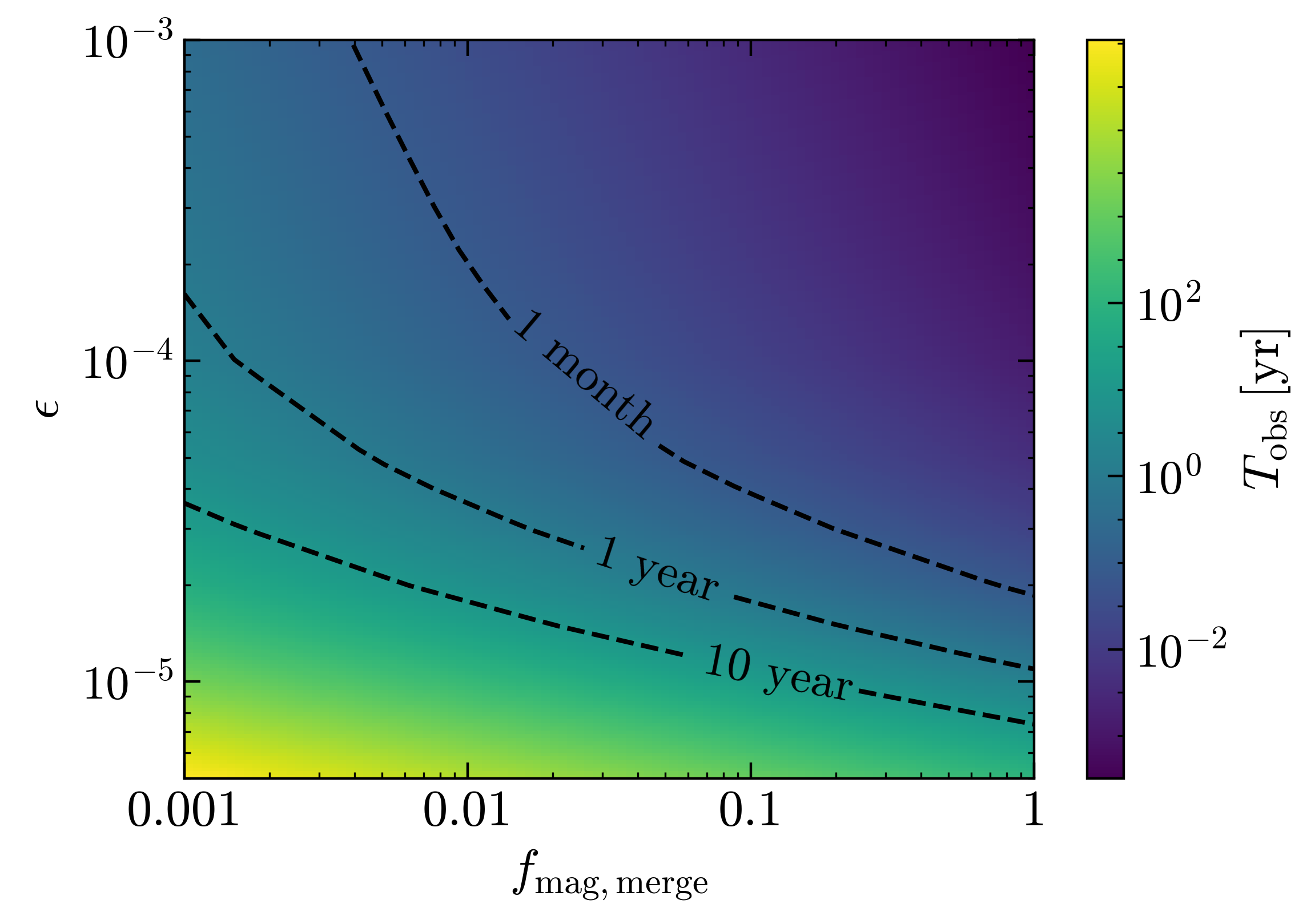}
\caption{The expected observing time for 3G detector network (CE+ET) to detect a positive event, as a function of $f_{\rm mag,merge}$ in NSBH binaries and the ellipticity $\epsilon$. We assume a rate of $\mathcal{R}_{\rm NSBH}={20}\,{\rm Gpc^{-3}\,yr^{-1}}$ for NSBH mergers, which agrees with the current estimate $9.1-84\,{\rm Gpc^{-3}\,yr^{-1}}$ from GW detections~\cite{LIGOScientific:2025pvj}.
}
\label{fig:fraction}
\end{figure}

We finally present a population framework for interpreting future detections or non-detections. 
For definiteness, we focus on NSBH systems, although it is also interesting to explore similar scenarios in BNS systems.
In the following, we take a reference rate of $\mathcal{R}_{\rm NSBH}=20\,{\rm Gpc^{-3}\,yr^{-1}}$ for NSBH mergers~\cite{LIGOScientific:2025pvj}.
Then the expected merger rate for a resonance-locking NSBH system is approximately
\begin{equation}
\begin{split}
    &\mathcal{R}_{\rm lock} = f_{\rm mag,merge}\mathcal{P}_{\rm lock}\mathcal{R}_{\rm NSBH} \\
    &\approx {0.12}\,f_{\rm mag,merge}\left(\frac{\epsilon}{10^{-5}}\right)^{1/2}\left(\frac{\mathcal{R}_{\rm NSBH}}{{20}\,{\rm Gpc^{-3}\,yr^{-1}}}\right){\rm Gpc^{-3}\,yr^{-1}}\,.
\end{split}
\end{equation}
Here $f_{\rm mag,merge}$ denotes the fraction of NSBH mergers in which the neutron star retains a sufficiently large magnetic ellipticity to produce a signal in the ground-based detector band.
This quantity is highly uncertain and is expected to be small. 
In Appendix~\ref{sec:fmag}, we provide a rough estimate to assess its sensitivity to both the short-delay tail of the tertiary-driven NSBH population and the assumed magnetic-field evolution.
The latter is modeled using illustrative field-decay tracks extracted from recent coupled core-crust magnetic-evolution simulations from Ref.~\cite{Skiathas:2024jah}.
Under our benchmark assumptions, $f_{\rm mag,merge}$ can reach $\sim10^{-2}$, making the resulting resonance-locking signal potentially relevant for CE+ET (see Fig.~\ref{fig:fraction}). More conservative assumptions, however, may substantially reduce this probability.

Assuming a redshift-dependent volumetric merger rate (number of mergers per comoving volume per cosmic time at redshift $z$) as $\mathcal{R}(z) =\mathcal{R}_{\rm lock}\times (1+z)^{3.2}e^{-z^2/3}$ for $z<6$, we obtain the resolvable merger events as
\begin{equation}
    \dot N_{\rm res} = \int \frac{dV_c(z)}{dz}\frac{\mathcal{R}(z)}{1+z}f_{\rm res}(z)dz\,,
\end{equation}
where $V_c(z)$ is the comoving volume up to redshift $z$. $f_{\rm res}(z)$ is the fraction of resonance-locking events that can be individually resolved by CE+ET, obtained from the synthetic population study (see Appendix~\ref{sec:NSBH pop}).

The expected observing time for CE+ET to detect one positive event reads
\begin{equation}
    T_{\rm obs} = \frac{1}{\dot N_{\rm res}}\,.
\end{equation}
In Fig.~\ref{fig:fraction} we show $T_{\rm obs}$ as a function of $f_{\rm mag}$ and $\epsilon$. 
We note that ET performs better than CE at low frequencies ($<5\,{\rm Hz}$), so for relatively small values of $\epsilon$ ($\lesssim10^{-5}$), the detection capability is determined primarily by ET.

\section{Discussion and conclusion}
\label{sec:discussion}
In this work, we present a novel resonance-locking mechanism--the sustained spin-orbit resonance--that can arise in compact-star binaries containing an intrinsically deformed compact object.
In this mechanism, the permanent quadrupole of the deformed object couples to the tidal field of its companion, allowing the stellar spin to lock to the orbital motion during the inspiral. 
The resonance locking and breaking produce a characteristic phase signature in the gravitational waveform, providing a source-agnostic framework for searching for intrinsic ellipticity in compact binaries.

We have discussed magnetar as one possible astrophysical realization for this mechanism.
Depending on the neutron star magnetic-field decay and merger-delay distribution, the resulting resonance-locking signal may enter the sensitivity band of future ground-based detectors, while decihertz detectors such as DECIGO may provide a more promising avenue.
We have also carried out a search for this locking signal in all GW events containing a neutron star up to O4a, finding no positive event so far.
Nevertheless, further searches should be conducted to facilitate discovery or place meaningful constraints.
Injection studies show that, when the locking signal lies within the sensitive band of the detector and accumulates sufficient signal-to-noise ratio, the breaking frequency, and hence the ellipticity, can be accurately reconstructed together with the stellar moment of inertia. Such measurements would provide new probes of neutron-star structure and the underlying dense-matter EOS~\cite{Lattimer:2000nx,Miao:2021gmf}.
Moreover, the simultaneous measurement of the moment of inertia and the tidal Love number in this system would enable a unique way to test gravity through the EOS-independent I-Love relation~\cite{Yagi:2013awa}.

Beyond the magnetar scenario, the resonance-locking mechanism can be applied to a broad class of compact-object systems.
For example, systems involving elastic materials, such as neutron star crusts, solid-quark cores~\cite{Alford:2000ze}, or even solid-quark stars~\cite{Xu:2003xe,Miao:2020cqj}, can sustain relatively large ellipticities.
Moreover, exotic compact objects have been proposed to deviate from Kerr black holes by supporting finite asymmetric deformations~\cite{Raposo:2018xkf,Bianchi:2020bxa,Herdeiro:2020kvf}. 
The resonance locking framework can be applied to test these black hole mimickers and to probe their deviations from Kerr geometry.

\acknowledgments
We thank Yi-Fan Wang for the help with gravitational-wave data processing and parameter inference and Yong Gao for helpful discussion regarding the precession dynamics of triaxially-deformed neutron stars. 
Z.-Q. M. is supported by the Postdoctoral Innovation Talent Support Program of CPSF (No. BX20240223) and the China Postdoctoral Science Foundation funded project (No. 2024M761948).
H. Y. is supported by the Natural Science Foundation of China (Grant 12573048).

This research has made use of data or software obtained from the Gravitational Wave Open Science Center (gwosc.org), a service of the LIGO Scientific Collaboration, the Virgo Collaboration, and KAGRA. This material is based upon work supported by NSF's LIGO Laboratory which is a major facility fully funded by the National Science Foundation, as well as the Science and Technology Facilities Council (STFC) of the United Kingdom, the Max-Planck-Society (MPS), and the State of Niedersachsen/Germany for support of the construction of Advanced LIGO and construction and operation of the GEO600 detector. Additional support for Advanced LIGO was provided by the Australian Research Council. Virgo is funded, through the European Gravitational Observatory (EGO), by the French Centre National de Recherche Scientifique (CNRS), the Italian Istituto Nazionale di Fisica Nucleare (INFN) and the Dutch Nikhef, with contributions by institutions from Belgium, Germany, Greece, Hungary, Ireland, Japan, Monaco, Poland, Portugal, Spain. KAGRA is supported by Ministry of Education, Culture, Sports, Science and Technology (MEXT), Japan Society for the Promotion of Science (JSPS) in Japan; National Research Foundation (NRF) and Ministry of Science and ICT (MSIT) in Korea; Academia Sinica (AS) and National Science and Technology Council (NSTC) in Taiwan.

\bibliography{references}

\appendix

\section{Eccenetric orbit}
\label{sec:ecc-orb}
In the main text, we discuss spin-orbit resonance locking for circular orbits.
In this section, we extend the analysis to eccentric orbits, working within a small-eccentricity expansion.
For simplicity, we still consider the case where the long axis of the neutron star lies within the orbital plane. In the next section, we will briefly discuss the more general scenario, where the long axis is oriented off the orbital plane.

\subsection{Equation of Motion and the Probability of Capturing Resonance}
The eccentric orbit can be described by
\begin{align}
    r &= a(1-e\cos u)\,,\\
    f & = 2\arctan\left[\left(\frac{1+e}{1-e}\right)^{1/2}\tan \frac{u}{2}\right]\,,\\
    \ell &\equiv n(t-t_0) = u-e\sin u\,,
\end{align}
where $a$ is the semi-major axis and $e$ is the eccentricity. $n=\sqrt{(M+M^\prime)/a^3}$ is the mean motion. The auxiliary variables $u$, $f$ and $\ell$ are the eccentric, true and mean anomalies, respectively. $t_0$ is a constant of integration.
With these notations, we can rewrite Eqs.~(2) and (3) in the main text as
\begin{align}
    &\ddot \ell =\dot n\,,\label{eq:ecc fddot}\\
    &\ddot \theta -\frac32\epsilon n^2\frac{q}{1+q}\frac{a^3}{r^3}\sin\left[2(f-\theta)\right] = 0\,,\label{eq:ecc thetaddot}
\end{align}
where we again omit the $I/I_{\rm orb}$ term and $\dot \Omega_{\rm spin}$ as done in the main text.
Redefining $\gamma = 2\theta-2p\ell$, we get 
\begin{equation}\label{eq:gamma}
    \ddot \gamma+3\epsilon n^2\frac{q}{1+q}\frac{a^3}{r^3}\sin(\gamma+2p\ell -2f) = -2p\dot n\,,
\end{equation}
By averaging all the terms in Eq.~(\ref{eq:gamma}) over one orbital period and expanding in terms of $e$, we then obtain~\cite{Murray:2000} 
\begin{equation}\label{eq:gamma average}
    \ddot \gamma +3\epsilon n^2\frac{q}{1+q}H(p,e)\sin \gamma =-2p\dot n\,,
\end{equation}
where $H(p,e)=G_{20(2p-2)}(e)$ with $G_{lpq}(e)$ the Kaula eccentricity function~\cite{Kaula:1966}.
Only values of $p$ that are integer multiples of $\frac{1}{2}$ will make $H(p,e)$ nonzero. For example, to $\mathcal{O}(e^4)$,
\begin{align}
    H(1,e) &= 1-\frac52e^2+\frac{13}{16}e^4\,,\\
    H(\frac{3}{2},e) &= \frac{7}{2}e-\frac{123}{116}e^3\,,\\
    H(2,e) &= \frac{17}{2}e^2-\frac{115}{6}e^4\,,\\
    H(\frac52,e) &= \frac{845}{48}e^3\,.
\end{align}

We rewrite Eq.~(\ref{eq:gamma average}) as 
\begin{equation}\label{eq:pendulum apx}
    \ddot \gamma +\mathcal{B}\sin \gamma+\mathcal{C} = 0\,,
\end{equation}
with 
\begin{align}
    \mathcal{B} &= 3\epsilon n^2\frac{q}{1+q}H(p,e)\,,\\
    \mathcal{C} &= 2p\dot n\,.
\end{align}
According to the general theory of resonance capture~\cite{Henrard:1982}, as $B_1=-2\pi\mathcal{C}-4\mathcal{\dot B}\mathcal{B}^{-1/2}$ is always negative for the neutron-star binaries in the inspiral phase, resonance can only occur if $B_1+B_2=-8\mathcal{\dot B}\mathcal{B}^{-1/2}<0$ ($B_1$ and $B_2$ are the two balance of energy defined in Ref.~\cite{Henrard:1982}).
This condition requires $\mathcal{\dot{B}}>0$, which holds only for the $p = 1$ and $p = 3/2$ cases. {\it Therefore, for eccentric neutron-star binaries, spin-orbit resonance locking can only occur in the 1:1 and 3:2 modes.}
The probability of capturing into resonance is then given by
\begin{equation}
\begin{split}
    &\mathcal{P}(p,e)= \frac{2}{1+\frac{\pi}{2}\frac{\mathcal{B}^{1/2}\mathcal{C}}{\mathcal{\dot B}}} \approx \frac{4}{\pi}\frac{\mathcal{\dot B}}{\mathcal{B}^{1/2}\mathcal{C}}\\  
    &=\frac{4\sqrt{3}}{\pi p}\left(\frac{q}{1+q}\right)^{1/2}\epsilon^{1/2}\left(\sqrt{H(p,e)}+\frac{\dot H(p,e)}{2\sqrt{H(p,e)}}\frac{n}{\dot n}\right)\,.
\end{split}
\end{equation}
Substituting the evolution for the semi-major axis $a$ and eccentricity $e$~\cite{Peters:1964zz},
\begin{align}
    \dot a &= -\frac{64}{5}\frac{MM^\prime (M+M^\prime)}{a^3}\frac{1}{(1-e^2)^{7/2}}\left(1+\frac{73}{24}e^2+\frac{37}{96}e^4\right)\,,\\
    \dot e &= -\frac{304}{15}\frac{MM^\prime (M+M^\prime)}{a^4}\frac{e}{(1-e^2)^{5/2}}\left(1+\frac{121}{304}e^2\right)\,,
\end{align}
we obtain,
\begin{align}
    \mathcal{P}(1,e)& = \frac{4\sqrt{3}}{\pi}\left(\frac{q}{1+q}\right)^{1/2}\epsilon^{1/2}\left(1+\frac{25}{18}e^2\right)\,,\\
    \mathcal{P}(\frac32,e) &= \frac{4\sqrt{3}}{\pi}\left(\frac{q}{1+q}\right)^{1/2}\epsilon^{1/2}\frac{17}{54}\sqrt{\frac{7}{2}}e^{1/2}\,.
\end{align}

\subsection{Gravitational waveform correction}
\label{subsec:waveform}
In the main text, we derive the gravitational-wave phase correction due to resonance locking for circular orbits, which is not applicable to eccentric orbits. 
Here, we provide an alternative correction for eccentric orbits, based on the post-Newtonian inspiral-only waveform model ({\sc{pyEFPE}}) developed in Ref.~\cite{Morras:2025nlp}.
The equations of motion for $y$ [$(1-e^2)y^2=(M_t\omega)^{2/3}$ with $M_t=M+M^\prime$ the total mass and $\omega$ the mean orbital angular velocity] and $e^2$ in Ref.~\cite{Morras:2025nlp} are given by
\begin{align}
    \mathcal{D}y &= \nu y^9\left(a_0+\sum_{n=2}^{6}a_ny^n\right)\,,\label{eq:Dy}\\
    \mathcal{D}e^2 &= -\nu y^8\left(b_0+\sum_{n=2}^{6}b_n y^n\right)\,,\label{eq:De}
\end{align}
where $\nu=\mu/M_t$ is the symmetric mass ratio with the reduced mass $\mu=MM^\prime/M_t$.
$a_n$ and $b_n$ are PN coefficients with 
\begin{align}
    a_0 &= \frac{32}{5}+\frac{28}{5}e^2\,,\\
    b_0 &= \frac{608}{15}e^2+\frac{242}{15}e^4\,,
\end{align}
and $\mathcal{D}$ is defined by
\begin{equation}
    \mathcal{D} = \frac{M_t}{(1-e^2)^{3/2}}\frac{d}{dt}\,.
\end{equation}
Adding resonance locking effect, these equations should be modified as
\begin{align}
    \mathcal{D}y &= \mathcal{D}^0y+\mathcal{D}^1 y\,,\\
    \mathcal{D}e^2 &= \mathcal{D}^0e^2+\mathcal{D}^1 e^2\,,
\end{align} 
where $\mathcal{D}^0$ terms are given by Eqs.~(\ref{eq:Dy}) and (\ref{eq:De}). 
Since the locking occurs at very low $y$, we consider only the Newtonian-order contribution to $\mathcal{D}^1$. 
At Newtonian order, the energy and angular momentum of a spin-orbit locked system are given by
\begin{align}
    E &= -\frac{\mu}{2}y^2(1-e^2)+\frac{1}{2}\frac{I}{M_t^2}y^6(1-e^2)^3\,,\label{eq:Etot}\\
    L &= \mu M_ty^{-1}+\frac{I}{M_t}y^3(1-e^2)^{3/2}\,,\label{eq:Ltot}
\end{align}
and the energy and angular momentum losses caused by gravitational wave radiation are~\cite{Peters:1964zz}
\begin{align}
    \frac{dE}{dt} &= -\frac{32}{5}\frac{\mu^2}{M_t^2}y^{10}(1-e^2)^{3/2}\left(1+\frac{73}{24}e^2+\frac{37}{96}e^4\right)\,,\label{eq:dot Etot}\\
    \frac{dL}{dt} &= -\frac{32}{5}\frac{\mu^2}{M_t}y^{7}(1-e^2)^{3/2}\left(1+\frac{7}{8}e^2\right)\,.\label{eq:dot Ltot}
\end{align}
Combining Eqs.~(\ref{eq:Etot}-\ref{eq:dot Ltot}), we get the correction as
\begin{align}
    D^{1}y &=\nu y^{13}c_0\frac{3(1-e^2)^{1/2}I}{\mu M_t^2-3(1-e^2)^2Iy^4}\,,\\
    D^{1}e^2 &= -\nu y^{12}c_0\frac{6(1-e^2)(1-e^2-\sqrt{1-e^2})I}{\mu M_t^2-3(1-e^2)^2Iy^4}\,,
\end{align}
where 
\begin{equation}
    c_0 =a_0(1-e^2)+\frac{b_0}{2}\,.
\end{equation}

\begin{figure}%[h]--------------------------------------------------------------------------
\centering
\includegraphics[width=2.8in]{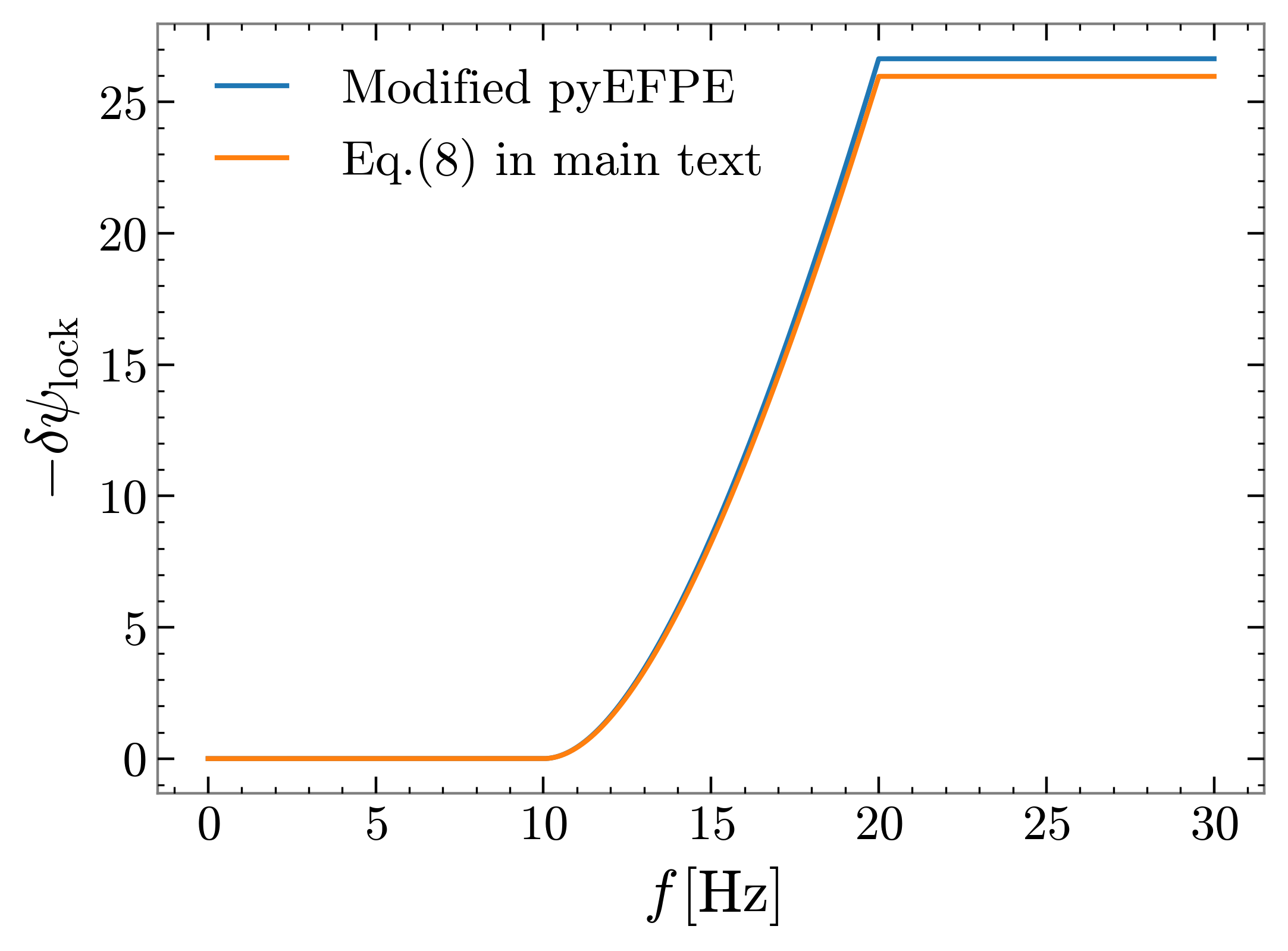}
\caption{Phase correction obtained from the modified {\sc pyEFPE} model for a non-eccentric binary merger, with $M=M^\prime=1.4\,M_\odot$. The locking start from $10\,{\rm Hz}$ and break at $f_{\rm br}=20\,{\rm Hz}$. The result is compared with the phase correction for circular orbits, Eq.~(\ref{eq:phase corr}), derived in the main text.
}
\label{fig:phase_err}
\end{figure}

We incorporated this correction to the equations of motion into the {\sc{pyEFPE}} waveform model. Figure~\ref{fig:phase_err} shows the phase correction obtained from the modified {\sc{pyEFPE}} model for a non-eccentric binary, compared with the phase correction for the circular orbit derived in the main text [cf. Eq.~(\ref{eq:phase corr})]. It can be seen that both corrections are consistent. When searching in GW200105 data, we used this modified {\sc{pyEFPE}} model.

\begin{figure}%[h]--------------------------------------------------------------------------
\centering
\includegraphics[width=2.8in]{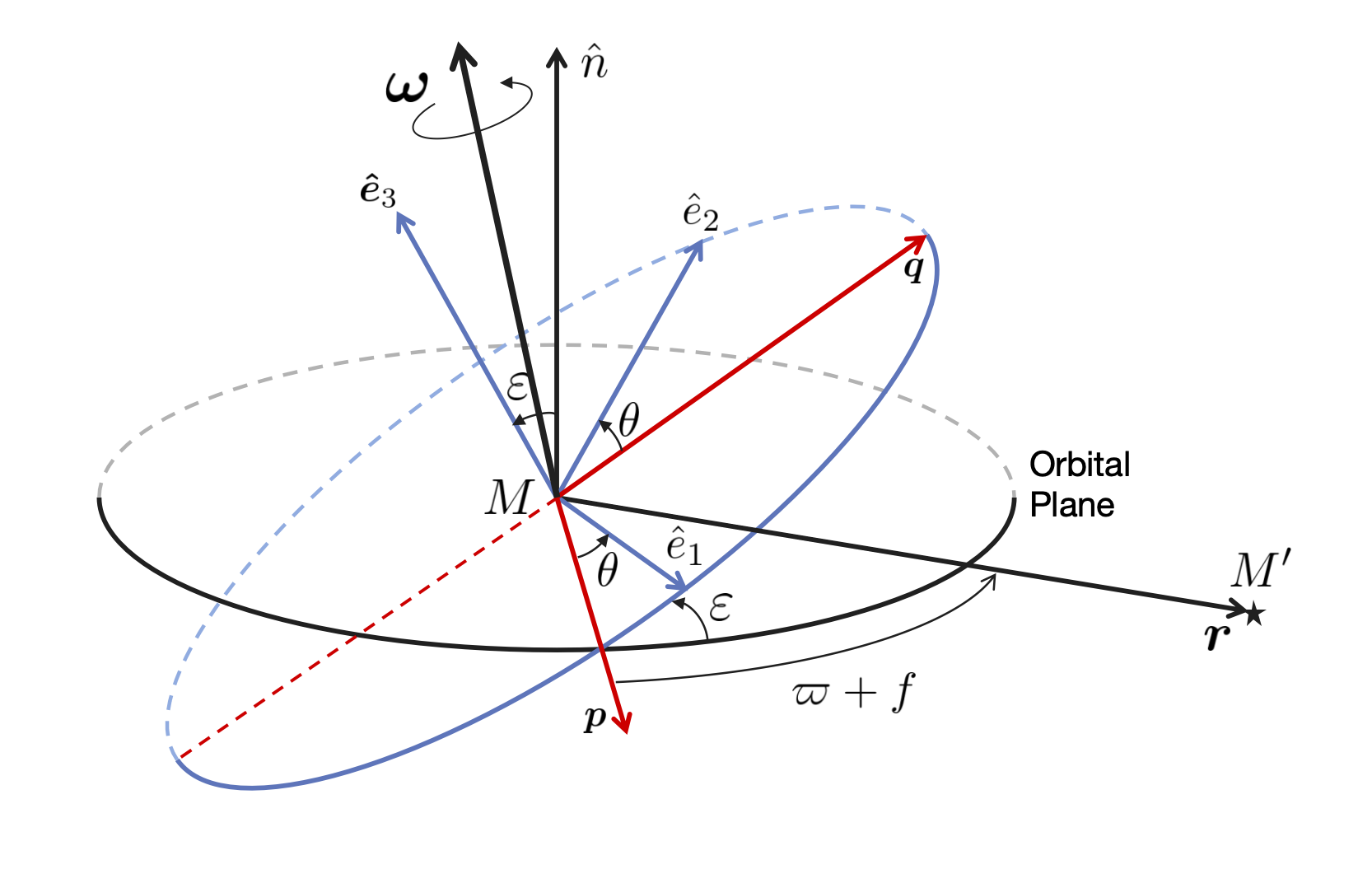}
\caption{The geometry and the motion of a deformed star $M$ in the body frame and orbital plane, with the companion $M^\prime$ being located at $\boldsymbol{r}$. $\boldsymbol{\omega}$ is the angular velocity.
$\boldsymbol{\hat e}_1$, $\boldsymbol{\hat e}_2$ and $\boldsymbol{\hat e}_3$ are the three unit eigenvectors along the principal axes of moment of inertia. $\boldsymbol{\hat n}$ is a unit vector normal to the orbital plane. 
$\boldsymbol{p}$ denotes the direction of equinox.
$\theta$ is the angle between $\boldsymbol{p}$ and $\boldsymbol{\hat e}_1$ and $\varpi+f$ is the angle between $\boldsymbol{p}$ and $\boldsymbol{r}$, where $f$ is the true anomaly and $\varpi$ is the longitude of periapsis. $\varepsilon$ is the angle between $\boldsymbol{\hat n}$ and $\boldsymbol{\hat e}_3$.
}
\label{fig:geometry}
\end{figure}

\section{General Case for Spin-Orbit Resonance}
\label{sec:misalign}
For a deformed neutron star, we denote the three unit eigenvectors along the principal axes as $\boldsymbol{\hat e}_1$, $\boldsymbol{\hat e}_2$ and $\boldsymbol{\hat e}_3$, with corresponding eigenvalues $I_1$, $I_2$ and $I_3$.
The angular velocity is $\boldsymbol{\omega}=\omega_1\boldsymbol{\hat e}_1+\omega_2\boldsymbol{\hat e}_2+\omega_3\boldsymbol{\hat e}_3$. 
We illustrate the geometry and the motion of the neutron star in Fig.~\ref{fig:geometry}.
In general, the spin vector $\boldsymbol{\omega}$ does not coincide with the principal axes. Additionally, $\boldsymbol{\omega}$ also does not need to align with the orbital normal $\boldsymbol{\hat n}$. These features result in more complex dynamics of the neutron star, including two types of precession: one is the precession of $\boldsymbol{\omega}$ about the normal to the orbital plane with rate $\Omega_{p,1}\sim \epsilon n^2/\omega$, and the other is the precession of $\boldsymbol{\hat e}_3$ about $\boldsymbol{\omega}$ with rate $\Omega_{p,2}\sim \epsilon \omega$. 
Near the resonance, we have $\omega \simeq n$, therefore both two precessions have a timescale of roughly $1/(\epsilon n)$, which is much longer than the resonance timescale $\sqrt{1/\dot n}$ for $\epsilon \lesssim 10^{-4}$.  
In the following discussion, we will neglect the precessions due to their relatively long timescales. A detailed study of the precessions may be pursued in future work.

We work in the body frame $(\boldsymbol{\hat e}_1,\boldsymbol{\hat e}_2,\boldsymbol{\hat e}_3)$. 
The Euler's equations of motion are 
\begin{align}
    I_1\dot \omega_1-(I_2-I_3)\omega_2\omega_3 = N_1\,,\\
    I_2\dot \omega_2-(I_3-I_1)\omega_3\omega_1 = N_2\,,\\
    I_3\dot \omega_3-(I_1-I_2)\omega_1\omega_2 = N_3\,,\label{eq:I3}
\end{align}
where the gravitational torques received by the companion are~\cite{Murray:2000}
\begin{align}
    N_1 = \frac{3M^\prime}{r^3}(I_3-I_2)(\boldsymbol{\hat r}\cdot \boldsymbol{\hat e}_2)(\boldsymbol{\hat r}\cdot \boldsymbol{\hat e}_3)\,,\label{eq:N1}\\
    N_2 = \frac{3M^\prime}{r^3}(I_1-I_3)(\boldsymbol{\hat r}\cdot \boldsymbol{\hat e}_3)(\boldsymbol{\hat r}\cdot \boldsymbol{\hat e}_1)\,,\label{eq:N2}\\
    N_3 = \frac{3M^\prime}{r^3}(I_2-I_1)(\boldsymbol{\hat r}\cdot \boldsymbol{\hat e}_1)(\boldsymbol{\hat r}\cdot \boldsymbol{\hat e}_2)\,.\label{eq:N3}
\end{align}
Note we have 
\begin{equation}\label{eq:angle}
 \left\{\!
\begin{aligned}
&\boldsymbol{\hat r}\cdot \boldsymbol{\hat e}_1=\cos w\cos \theta+\sin w\sin\theta\cos\varepsilon\,,\\ 
&\boldsymbol{\hat r}\cdot \boldsymbol{\hat e}_2=-\cos w\sin\theta+\sin w\cos\theta\cos\varepsilon\,,\\
&\boldsymbol{\hat r}\cdot \boldsymbol{\hat e}_3=-\sin w\sin\varepsilon\,,
\end{aligned}
\right.
\end{equation}
where $w=\varpi+f$.
Plugging Eq.~(\ref{eq:angle}) into Eqs.~(\ref{eq:N1}-\ref{eq:N3}) we get 
\begin{align}
    N_1 = &\frac{3M^\prime}{2r^3}(I_3-I_2)\Big[\sin(\theta-2w)\cos^2\frac{\varepsilon}{2}\sin\varepsilon\nonumber\\
    &-\sin(\theta+2w)\sin^2\frac{\varepsilon}{2}\sin\varepsilon-\frac{1}{2}\sin\theta\sin2\varepsilon\Big] \,.\\
    N_2 = &\frac{3M^\prime}{2r^3}(I_1-I_3)\Big[\cos(\theta-2w)\cos^2\frac{\varepsilon}{2}\sin\varepsilon\nonumber\\
    &+\cos(\theta+2w)\sin^2\frac{\varepsilon}{2}\sin\varepsilon-\frac{1}{2}\cos\theta\sin2\varepsilon\Big] \,.\\
    N_3 = &-\frac{3M^\prime}{2r^3}(I_2-I_1)\Big[\sin(2\theta-2w)\cos^4\frac{\varepsilon}{2}\nonumber\\
    &+\sin(2\theta+2w)\sin^4\frac{\varepsilon}{2}+\frac{1}{2}\sin2\theta \sin^2\varepsilon\Big] \,.
\end{align}
We now consider that the rotation frequency $\omega_3=\dot \theta$ and the mean motion $n=\dot \ell$ are close to resonance ($\omega_3 \simeq pn$).
By doing average over one orbital period and retaining all the terms with argument $2(\theta-p\ell)$, we then obtain the average torque~\cite{Correia:2009pb}
\begin{equation}\label{eq:N3bar}
\begin{split}
    \langle N_3\rangle = &-\frac{3M^\prime}{8a^3}(I_2-I_1)\Big[(x+1)^2H(p,e)\sin2(\theta-p\ell-\varpi)\\
    &+(x-1)^2H(-p,e)\sin2(\theta-p\ell+\varpi)\\
    &+2(1-x^2)G(p,e)\sin2(\theta-p\ell)\Big]\,,
\end{split}
\end{equation}
where $x=\cos\varepsilon$ and $G(p,e)$ is a power series in $e$. For $p=1$ and $p=3/2$, we have 
\begin{equation}
    G(1,e) = \frac{9}{4}e^2+\frac{7}{4}e^4\quad {\rm and}\quad 
    G(\frac{3}{2},e) = \frac{53}{16}e^3\,.
\end{equation}
Substituting Eq.~(\ref{eq:N3bar}) into Eq.~(\ref{eq:I3}), we get
\begin{equation}\label{eq:omegadot}
    \dot\omega_3-\epsilon\omega_1\omega_2 = -\frac{\mathcal{B}_x}{2}\sin2(\theta-p\ell-\varpi_x)\,,
\end{equation}
where 
\begin{align}
    \left(\frac{\mathcal{B}_x}{\mathcal{B}}\right)^2= &4\big[(g+h^++h^-)^2-4g(h^++h^-)\sin^2\varpi\nonumber\\
    &-4h^+h^-\sin^22\varpi\big]\,,\\
    \tan2\varpi_x = &\frac{(h^+-h^-)\sin2\varpi}{g+(h^++h^-)\cos2\varpi}\,,
\end{align}
with 
\begin{align}
    \mathcal{B} = &3\epsilon n^2\frac{q}{1+q}\,,\\
    g =& \frac{1}{4}(1-x^2)G(p,e)\,,\\
    h^\pm=& \frac{1}{8}(x\pm1)^2H(\pm p,e)\,.
\end{align}

Let $\gamma = 2(\theta-p\ell -\varpi_x)$, then $\ddot \gamma = 2(\dot \omega_3-p\dot n)$, where we have neglect the slow variation of $\varpi_x$ caused by the precessions. Thus, Eq.~(\ref{eq:omegadot}) can be rewritten as
\begin{equation}
    \ddot \gamma+\mathcal{B}_x\sin\gamma = -2p\dot n+2\epsilon\omega_1\omega_2\,,
\end{equation}
which is the same as Eq.~(\ref{eq:pendulum}) or Eq.~(\ref{eq:pendulum apx}). We may further use Eq.~(\ref{eq:probability}) to evaluate the probability of capturing into resonance. 
Note here $\omega_1$ and $\omega_2$ can be considered as constants due to the long period of precession. 
For example, for a circular orbit with $e=0$, we have $g=h^-=0$ and thus 
\begin{equation}
    \mathcal{B}_x = \frac{3}{4}\epsilon n^2\frac{q}{1+q}(1+x)^2.
\end{equation}
If we further assume $\epsilon\omega_1\omega_2\ll \dot n$, we can get the probability as
\begin{equation}
    \mathcal{P}(1,0) = \frac{4\sqrt{3}}{\pi}\left(\frac{q}{1+q}\right)^{1/2}\epsilon^{1/2}\frac{1+x}{2}\,,
\end{equation}
which is reduced by a factor of $(1+x)/2$ compared to Eq.~(\ref{eq:probability}).

\section{Injection studies}
\label{sec:Injection}
In this section, we provide the details of the injection study. As stated in Sec.~\ref{sec:magnetar}, we consider two representative systems: a NSBH system with $M^\prime=5M=7\,M_\odot$ and an equal-mass BNS system with $M^\prime=M=1.4\,M_\odot$. Both the two systems are placed at a luminosity distance of $d_L=100\,{\rm Mpc}$ and the sky location of GW170817 with ${\rm R.A.}=3.446$ and ${\rm DEC.}=-0.408$~\cite{LIGOScientific:2017ync}.
We also assume a nonspinning companion and an orientation with $\theta_{\rm JN}=2.57$. 
Because the locked neutron star has a spin frequency below $f_{\rm br}$, we also neglect its spin in the waveform model.
The {\sc TaylorF2} waveform~\cite{Buonanno:2009zt}, including tidal effects~\cite{Vines:2011ud}, is used as the baseline waveform for the BNS system, 
while {\sc IMRPhenomXPHM} waveform~\cite{Pratten:2020ceb} is employed for the NSBH system (tidal effects are not included, as they are marginally small in this case). 

For each merger, we calculate the expected SNR as $\rho =\sqrt{\sum_{{\rm det.}\,k}\langle h_k|h_k\rangle}$, with the inner product
\begin{equation}
    \langle h_k|g_k\rangle = 2\int \frac{h_k^*(f)g_k(f)+h_k(f)g_k^*(f)}{S_{{\rm n},k}(f)}df\,,
\end{equation}
where $S_{{\rm n},k}$ and $h_k$ are the noise power spectrum density (PSD) and the signal strain of the $k$-th detector, respectively.
For the NSBH system, the SNRs can reach 117 with A+ and 2135 with CE+ET, while for the BNS case, the corresponding SNRs are 69 and 1190.
Owing to these high SNR values, we adopt the approach of injecting into zero-noise~\cite{Rodriguez:2013oaa}.
Parameter estimation is performed with {\sc Bilby}~\cite{Ashton:2018jfp} using the {\sc pymultinest} nested sampler~\cite{Buchner:2014nha}. 
We adopt $n_{\rm live}=2000$ and otherwise use the default settings implemented in {\sc Bilby}.
Injection and template waveforms are generated with LALSuite~\cite{lalsuite}, 
with both injection and recovery employing a $5\,{\rm Hz}$ low-frequency cutoff and terminating at the innermost stable circular orbit.

\begin{figure*}%[h]--------------------------------------------------------------------------
\centering
\includegraphics[width=3.2in]{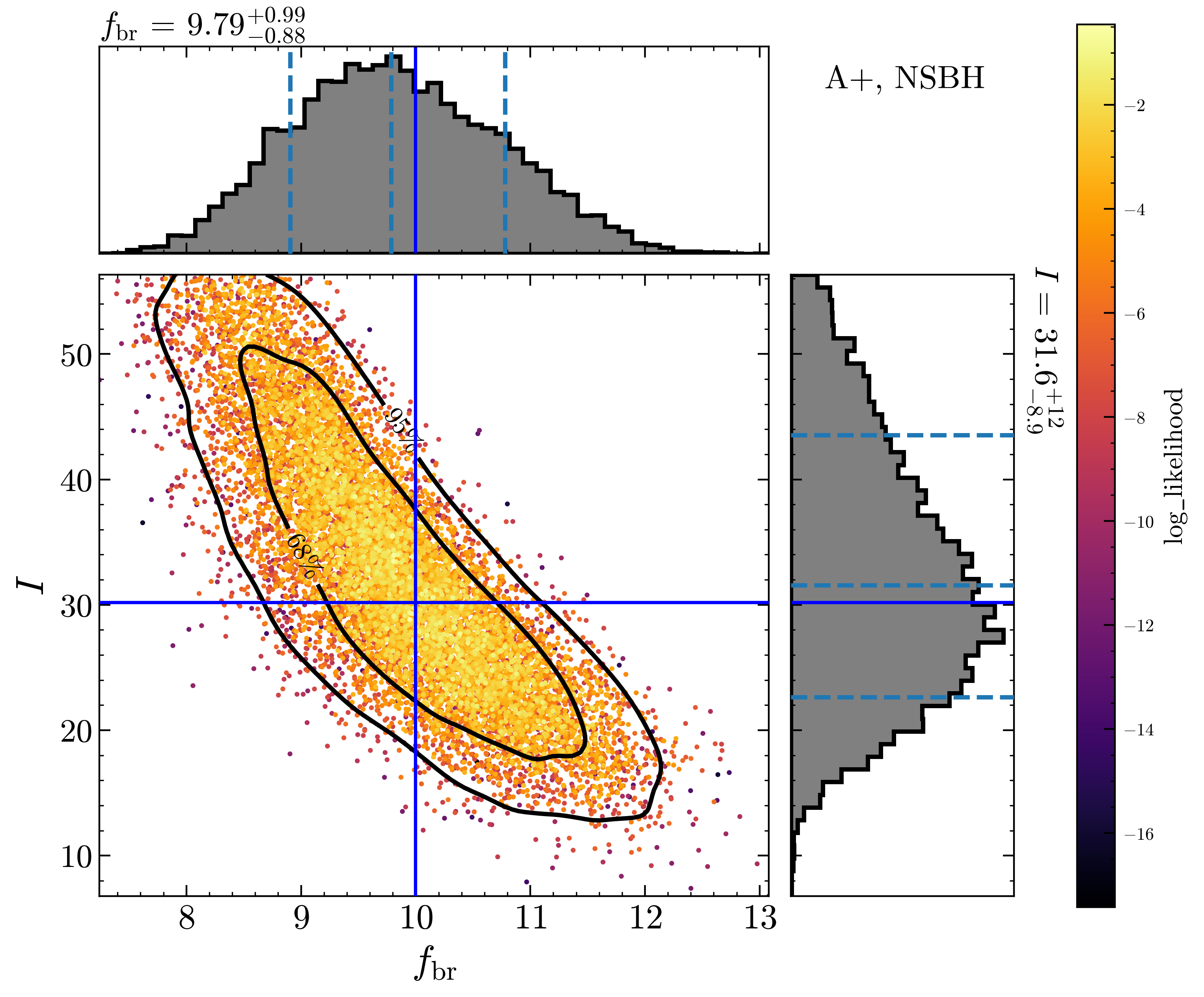}
\includegraphics[width=3.2in]{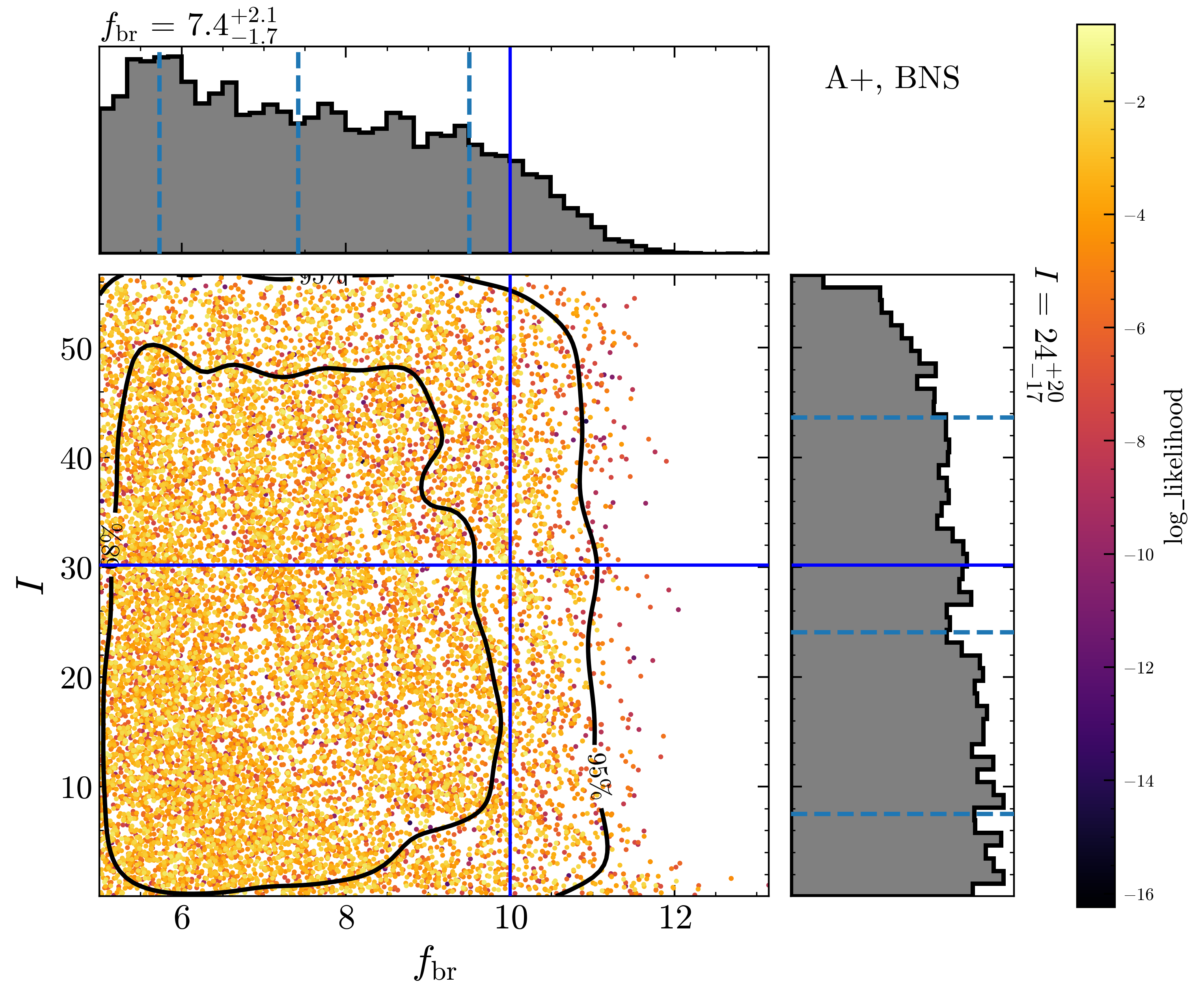}
\includegraphics[width=3.2in]{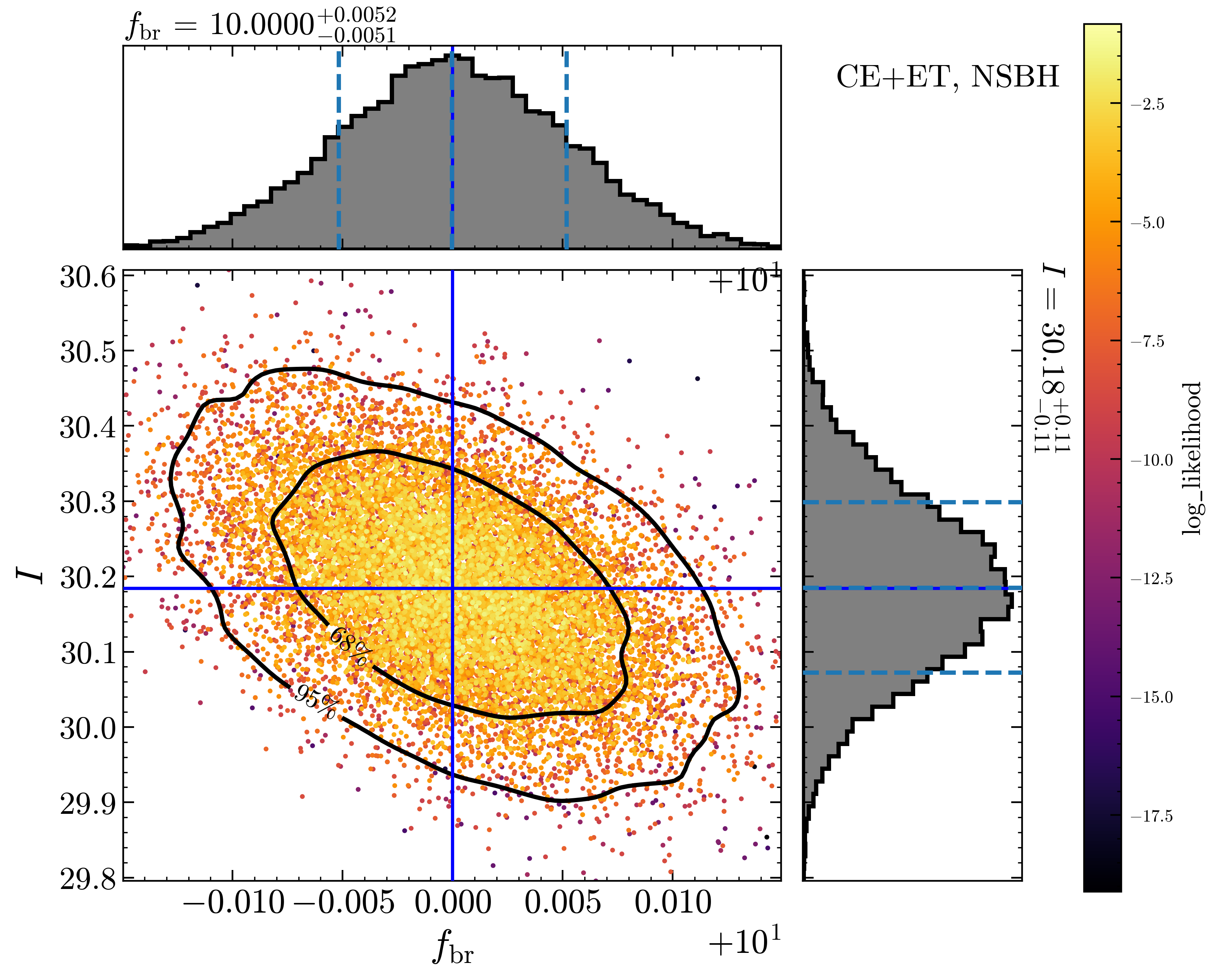}
\includegraphics[width=3.2in]{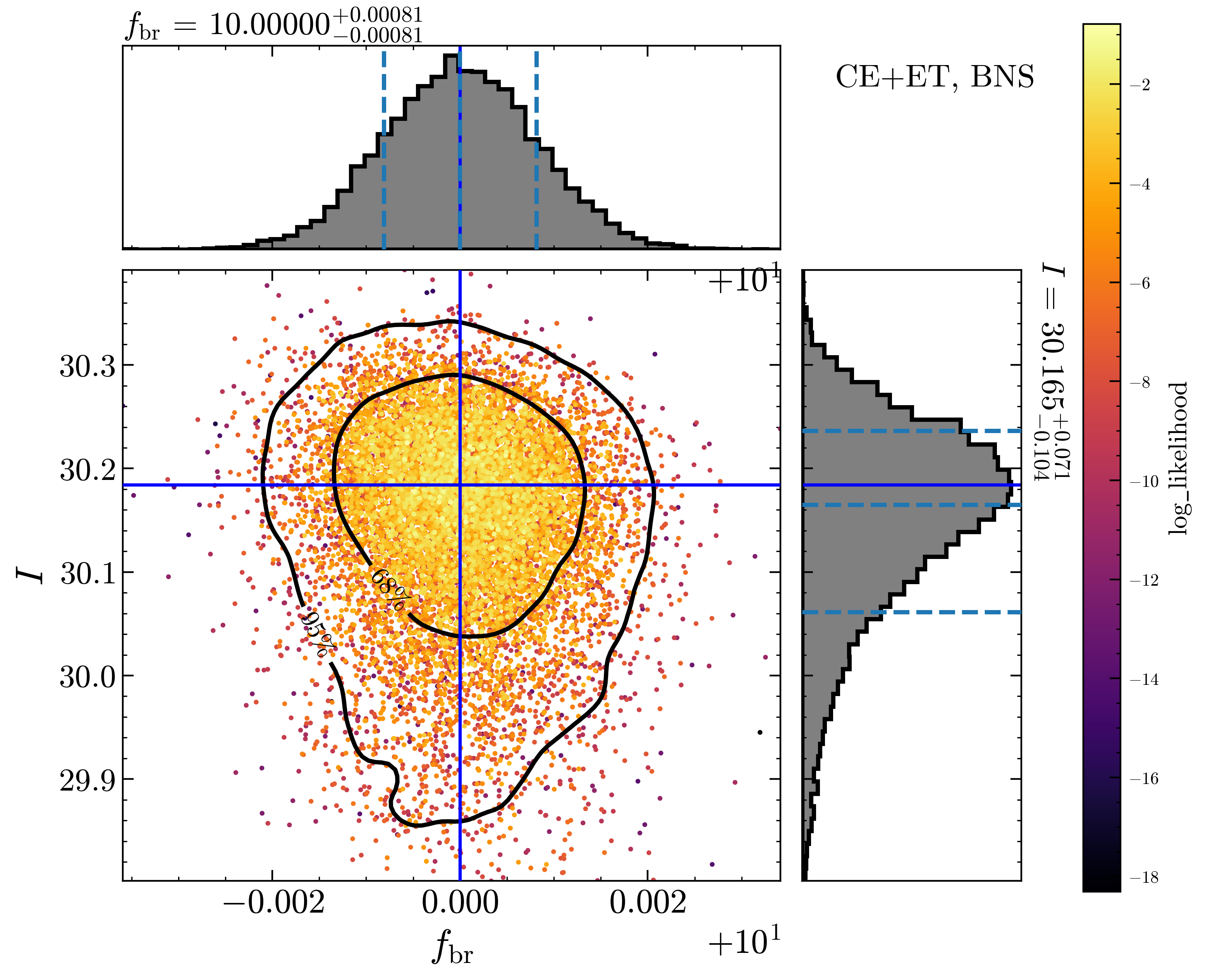}
\caption{Same as Fig.~\ref{fig:Inject} in the main text, but for injections with $f_{\rm br}=10\,{\rm Hz}$.
}
\label{fig:Inject10}
\end{figure*}

In addition to the $f_{\rm br}=20\,\rm{Hz}$ injections shown in the main text, we also examine injections with a lower breaking frequency of $f_{\rm br}=10\,\rm{Hz}$.
Figure~\ref{fig:Inject10} presents the posterior distributions of $f_{\rm br}$ and the moment of inertia $I$ for this case.
The figure shows that A+ either fails to recover the injected values or does so poorly, whereas the CE+ET network recovers the values accurately.

\section{Searches}
\label{sec:searches}
In this section, we present the details of the search studies. 
As stated in the main text, we have used publicly available GW strain data~\cite{LIGOScientific:2025snk} and noise PSDs and followed an inference procedure similar to that conducted by LVK collaboration.
For BNS mergers, we adopt the {\sc TaylorF2} waveform as the baseline waveform, while for the NSBH mergers we employ the {\sc IMRPhenomNSBH} waveform~\cite{Thompson:2020nei} except for GW200105.  
GW200105 has been reported with orbital eccentricity~\cite{Morras:2025xfu}, and therefore our circular-orbit GW phase correction, Eq.~(\ref{eq:phase corr}) in the main text, is not applicable in this case. Instead, we employ the modified {\sc pyEFPE} waveform as discussed above.
For GW190814, we adopt the {\sc IMRPhenomXPHM} waveform~\cite{Pratten:2020ceb}, which incorporates higher-order multipoles, as the event exhibits strong evidence for them~\cite{LIGOScientific:2020zkf}.
It should be note that the phase correction derived in the main text only applies to the $(\ell, m)=(2,2)$ multipole. For higher-order multipoles, we take
\begin{equation}\label{eq:phase corr HM}
    \delta\psi_{\ell m}(f) =\frac{m}{2}\delta\psi_{22}(\frac{2}{m}f)\,,
\end{equation}
where $\delta\psi_{22}(f)$ is given by Eq.~(\ref{eq:phase corr}) in the main text. We include $(2,2)$, $(3,3)$, $(2,1)$ and $(4,4)$ modes in the analysis. 
The phase correction is applied directly to the LALSuite-generated waveform in the frequency domain. 
For GW190814, which exhibit very weak precession ($\chi_p < 0.07$), this provides a good approximation.
During the search, we set a uniform prior for the moment of inertia ($I$), but we also impose a physical constraint of $I/M^3>4$,  where the value of 4 corresponds to the Schwarzschild black hole limit.

\begin{figure}%[h]--------------------------------------------------------------------------
\centering
\includegraphics[width=3.0in]{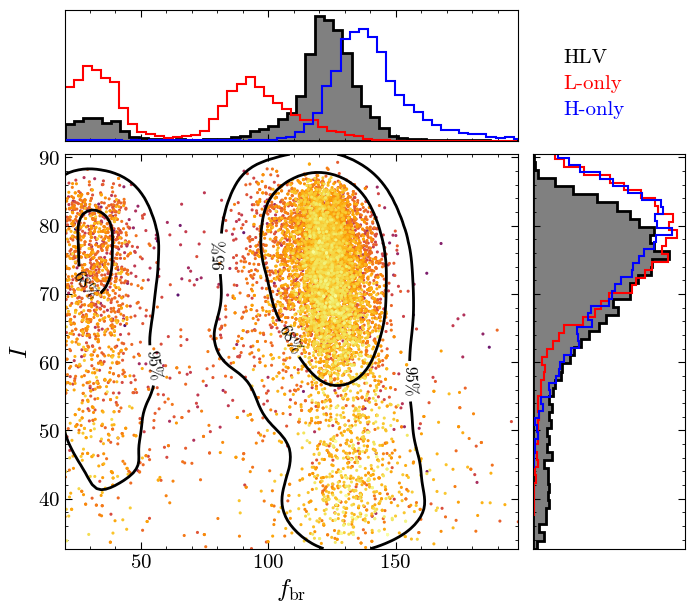}
\caption{Posterior distributions of breaking frequency $f_{\rm br}$ and moment of inertia $I$ with HLV data (black, grey shading), Livingston data only (red) and Hanford data only (blue), for GW190814.
}
\label{fig:GW190814}
\end{figure}

\begin{figure*}%[h]--------------------------------------------------------------------------
\centering
\includegraphics[width=7in]{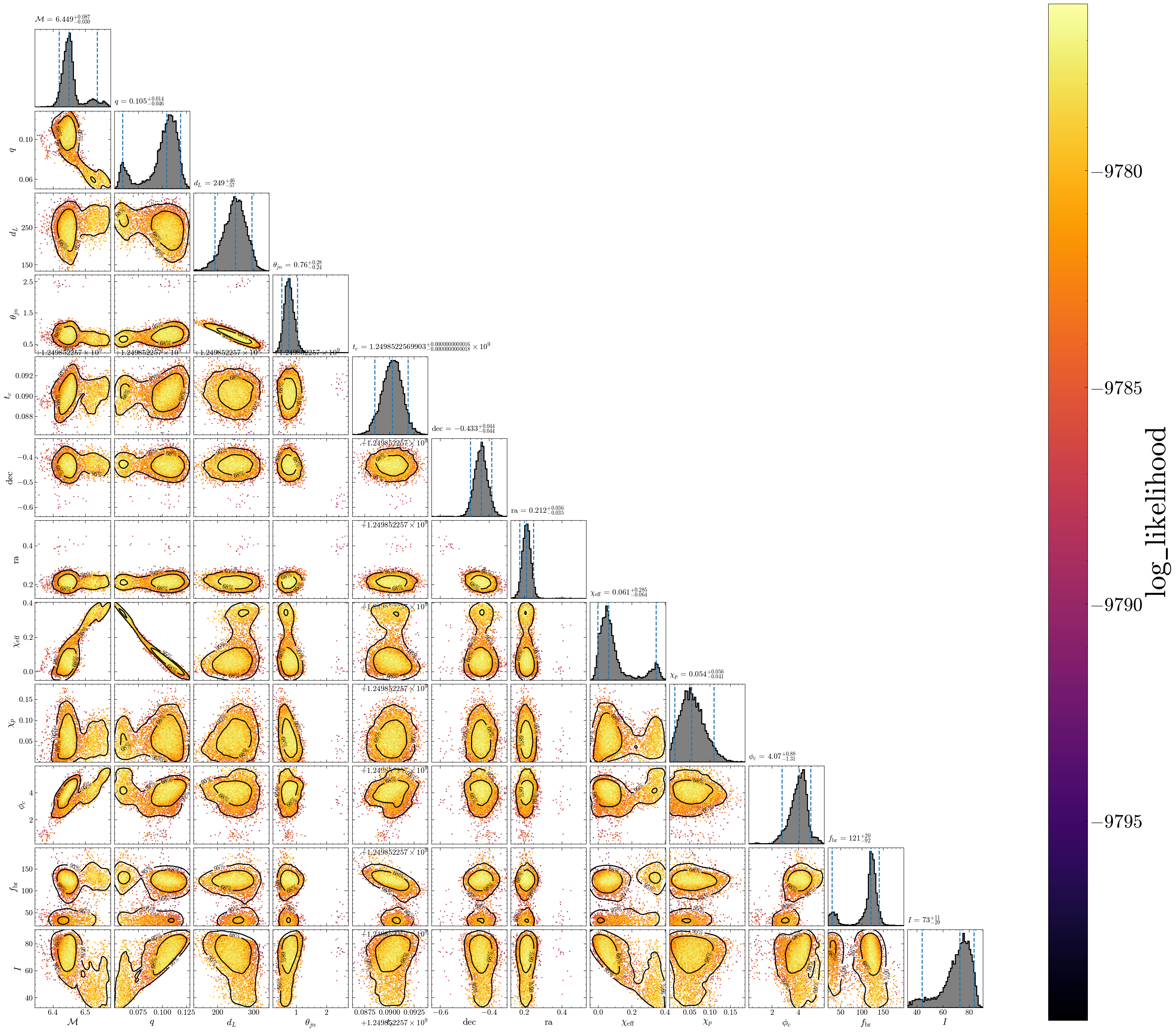}
\caption{The posterior distribution of all parameters in the search of resonance locking presented in Fig.~\ref{fig:GW190814} in the main text with data from GW190814.}
\label{fig:GW190814full}
\end{figure*}

In Table~\ref{tab:search_events}, we summarize the events analyzed in this work and the corresponding search configurations, including the low-frequency cutoff $f_{\rm low}$, the waveform model, and the resulting Bayes factor $\ln\mathcal{B}_0^1$.
In Fig.~\ref{fig:GW190814full}, we plot the full posterior distribution of parameters for the interesting mas-gap event GW190814 with Hanford-Livingston-Virgo data. 
We also examine the Hanford and Livingston data separately, and the results are plotted in Fig.~\ref{fig:GW190814}. 
In Hanford alone, a single peak appears near $f_{\rm br} \simeq 130\,{\rm Hz}$, while in Livingston alone, the posterior still exhibits a bimodal distribution for $f_{\rm br}$.
Note that LIGO Hanford was not in observing mode during the event, but its data is still usable~\cite{LIGOScientific:2020zkf}.
Furthermore, thunderstorms near Livingston caused noise in the $\lesssim 30\,{\rm Hz}$ band, as reported in~\cite{LIGOScientific:2020zkf}.

\begin{figure}%[h]--------------------------------------------------------------------------
\centering
\includegraphics[width=3.2in]{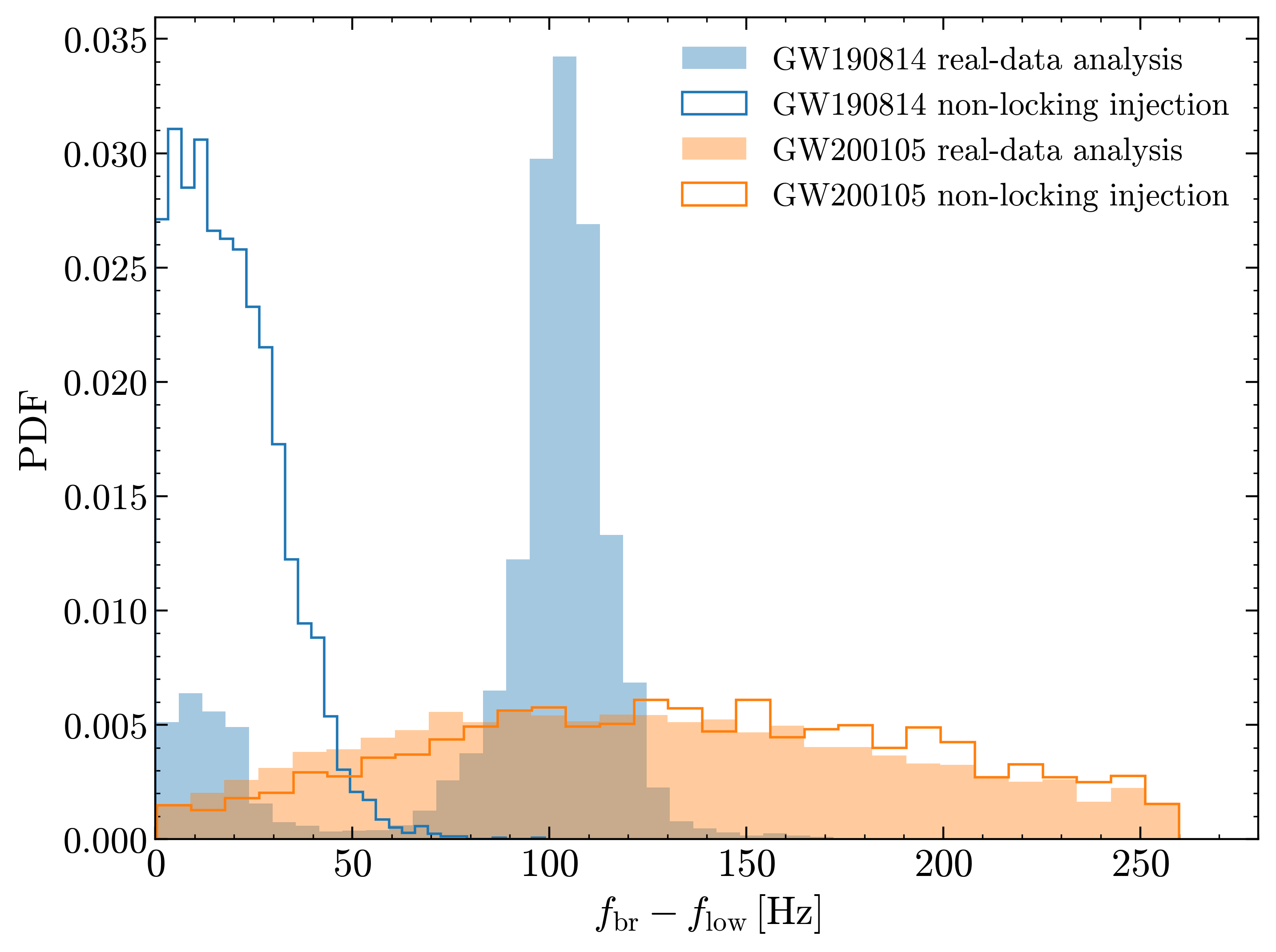}
\caption{{One-dimensional posterior distributions of $f_{\rm br}-f_{\rm low}$ inferred from Bayesian analysis with the resonance-locking corrected waveform model, applied to real data and to null injections, for GW190814 and GW200105.}
}
\label{fig:null_inj}
\end{figure}

{We also perform null-injection tests using waveforms without the resonance-locking phase correction for GW190814 and GW200105. 
The injected signals adopt the maximum a posteriori (MAP) parameters the two events, and are analyzed using the same inference setup as in the real-data analysis. The resulted posterior distributions of $f_{\rm br}-f_{\rm low}$ are shown in Fig.~\ref{fig:null_inj}.
For GW190814, the recovered posterior of $f_{\rm br}-f_{\rm low}$ collapses to the boundary ($f_{\rm br}-f_{\rm low}=0$), indicating that the sampler and waveform implementation do not artificially generate nonzero breaking frequencies. 
This result suggests that the secondary peak near $f_{\rm br}\sim120\,{\rm Hz}$ observed in the real-data analysis is driven by features in the data rather than by a bias in the inference procedure. 
For GW200105, the posterior remains broader and does not strongly accumulate at the boundary, reflecting the weaker constraints on this additional phase parameter given the detector sensitivity for this event.}
\begin{table*}[t]
\caption{{Summary of the analyzed events and the corresponding search configurations. 
For each event we list the prior of $f_{\rm br}$ used in the analysis, the base waveform model, and the resulting Bayes factor.}}
\begin{ruledtabular}
\begin{tabular}{lccc}
Event & Prior of $f_{\rm br}$ [Hz] & Waveform model& $\ln \mathcal{B}_0^1$  \\ \hline
GW170817 & $\mathcal{U}(23,100)$ & {\sc{TaylorF2}}+Eq.~(\ref{eq:phase corr}) & -1.46 \\
GW190425 & $\mathcal{U}(20,100)$ & {\sc{TaylorF2}}+Eq.~(\ref{eq:phase corr}) & -2.99 \\
GW200105 & $\mathcal{U}(20,280)$ & Modified pyEFPE & -1.70 \\
GW200115 & $\mathcal{U}(20,100)$ & {\sc{IMRPhenomNSBH}}+Eq.~(\ref{eq:phase corr}) & -0.90 \\
GW230518 & $\mathcal{U}(20,100)$ & {\sc{IMRPhenomNSBH}}+Eq.~(\ref{eq:phase corr}) & -0.59\\
GW230529 & $\mathcal{U}(20,100)$ & {\sc{IMRPhenomNSBH}}+Eq.~(\ref{eq:phase corr}) & -1.43 \\
GW190814 & $\mathcal{U}(20,180)$ & {\sc{IMRPhenomXPHM}}+Eq.~(\ref{eq:phase corr HM}) & -0.41 \\
\end{tabular}
\end{ruledtabular}
\label{tab:search_events}
\end{table*}

\section{NSBH population study}
\label{sec:NSBH pop}
In order to determine how much resonance-locking events that can be individually resolved by CE+ET, $f_{\rm res}(z)$, we performed a population analysis. We begin our calculation with the population model. For the black hole mass, we take a power-law distribution with a spectral index of $2.7$, featuring a sharp cutoff at the lower end of $4\,M_\odot$ and an upper bound of $40\,M_\odot$:
\begin{equation}
    p(m_{\rm BH}) \propto \left\{\!
\begin{aligned}
&m_{\rm BH}^{-2.7}\,, & 4\,M_\odot<m_{\rm BH}<40\,M_\odot\,,\\ 
&0\,, & {\rm otherwise}\,.
\end{aligned}
\right.
\end{equation}
For the neutron star mass, we assume a uniform mass distribution between $1.1\,M_\odot$ and $2.1\,M_\odot$, i.e., $p(m_{\rm NS})\sim {\rm U}(1.1\,M_\odot, 2.1\,M_\odot)$. 
For simplicity, we do not consider black hole spin in this analysis.
We then take a red-shift dependent volumetric NSBH merger rate with locking signal as 
\begin{equation}
    \mathcal{R}(z) = \mathcal{R}_{\rm lock}\times (1+z)^{3.2}e^{-z^2/3}\,,
\end{equation}
where the exponent $3.2$ is taken from the redshift dependence for the BBH merger rate inferred in GWTC-4~\cite{LIGOScientific:2025pvj}, and the factor $e^{-z^2/3}$ accounts for the suppression at high redshifts. We assume a flat $\Lambda$CDM cosmology with $H_0 = 67.4\,{\rm km\,s^{-1}\,Mpc^{-1}}$ and $\Omega_m=0.315$~\cite{Planck:2018vyg}.

\begin{figure}%[h]--------------------------------------------------------------------------
\centering
\includegraphics[width=3.2in]{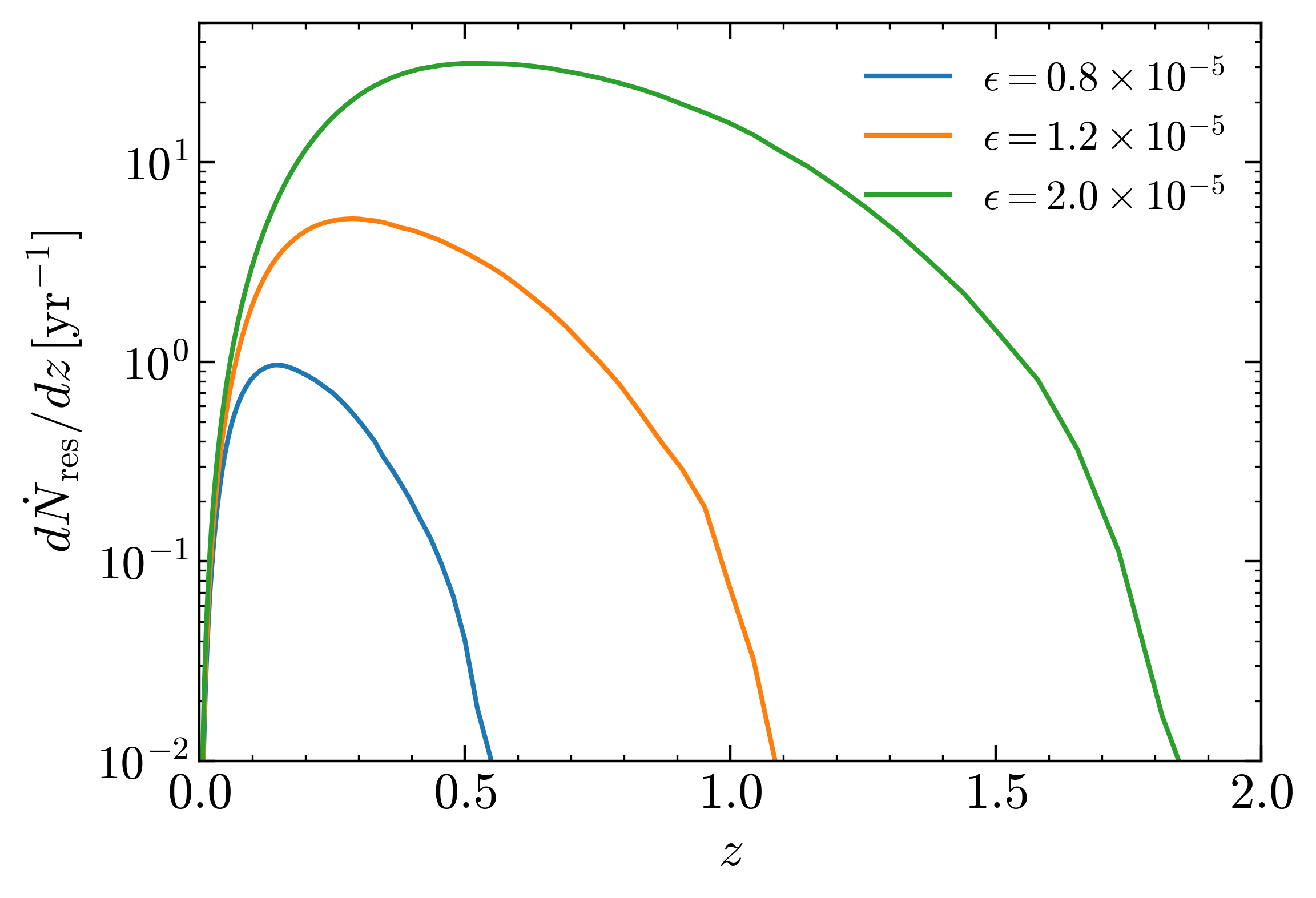}
\caption{Detection rate as a function of redshift $z$ for different $\epsilon$. The results shown in the figure are assuming a magnetar fraction of $f_{\rm mag}=1$ in NSBH binaries.}
\label{fig:dNdz}
\end{figure}

For each choice of $\epsilon$, we evaluate the corresponding breaking frequency $f_{\rm br}$ for the population, and assume locking signal can be resolved by CE+ET if the accumulated SNR in the frequency band $[1\,{\rm Hz}, f_{\rm br}]$ exceeds the threshold $\rho_{\rm thr}= 10$.
This allows us to calculate the fraction of resolvable events as a function of redshift, i.e., $f_{\rm res}(z)$. Then we get the differential resolvable merger events as
\begin{equation}
    d\dot N_{\rm res} =  f_{\rm res}(z)\frac{\mathcal{R}(z)}{1+z}dV_c(z)\,.
\end{equation}
In Fig.~\ref{fig:dNdz}, we show ${d{\dot N_{\rm res}}}/{dz}$ as a function of redshift $z$ for different values of $\epsilon$.

\section{A rough estimate of $f_{\rm mag,merge}$}
\label{sec:fmag}
In this section we provide a rough estimate of $f_{\rm mag,merge}$. 
This estimate is not intended as a robust population prediction. Rather, it is used to assess how $f_{\rm mag,merge}$ depends on uncertain magnetic-field decay and binary-evolution assumptions.
We approximate this fraction as
\begin{equation}
    f_{\rm mag,merge} = f_{\rm mag,0}(>B_{\rm int,0})P(f_{\rm br}>f_{\rm det,th})\,,
\end{equation}
where $P(f_{\rm br}>f_{\rm det,th})$ is the probability that, after magnetic-field decay over the merger delay time, the resulting breaking frequency lies in the detector band. 
The factor $f_{\rm mag,0}(>B_{\rm int,0})$ denotes the birth fraction of neutron stars whose initial internal fields exceed the adopted value $B_{\rm int,0}$.

Theoretical arguments and numerical simulations suggest that core-collapse supernovae can produce neutron stars with internal fields of order $B_{\rm int,0}\sim10^{15}\,{\rm G}$ or larger~\cite{Duncan:1992hi,Raynaud:2020ist}.
However, the birth fraction of neutron stars with such strong internal fields is poorly known. 
Observational estimates of the magnetar birth fraction based on galactic magnetar populations span a broad range of $\sim 0.01$--$0.5$~\cite{Gill:2007ra,Gullon:2015zca,Beniamini:2019bga,Pardo-Araujo:2026eby}, and should be regarded as an upper-limit guide for $f_{\rm mag,0}(>B_{\rm int,0})$.
Based on these considerations, we adopt $B_{\rm int,0}=10^{15}\,{\rm G}$ and $f_{\rm mag,0}=0.1$ as a benchmark; the actual fraction could be well below this value.

The evolution of the internal magnetic field is highly complex, and a robust calculation of $B_{\rm int}(t)$ would require solving the coupled thermal and magnetic-field evolution with detailed microphysics~\cite{Pons:2025qzc}. 
Such a calculation is beyond the scope of this work.
Instead, we adopt the numerical magnetic-field evolution tracks of Skiathas and Gourgouliatos~\cite{Skiathas:2024jah} as illustrative field-evolution models. Their simulations follow the coupled magnetic evolution of the neutron-star crust and core, including Hall drift and Ohmic dissipation in the crust, and ambipolar diffusion in the core. 
They also consider four representative initial magnetic-field configurations: purely poloidal, purely toroidal, mixed poloidal--toroidal, and ``twisted-torus'' fields.
We use all four cases to assess the sensitivity of our results to the uncertain initial magnetic geometry.

\begin{figure}%[h]--------------------------------------------------------------------------
\centering
\includegraphics[width=2.8in]{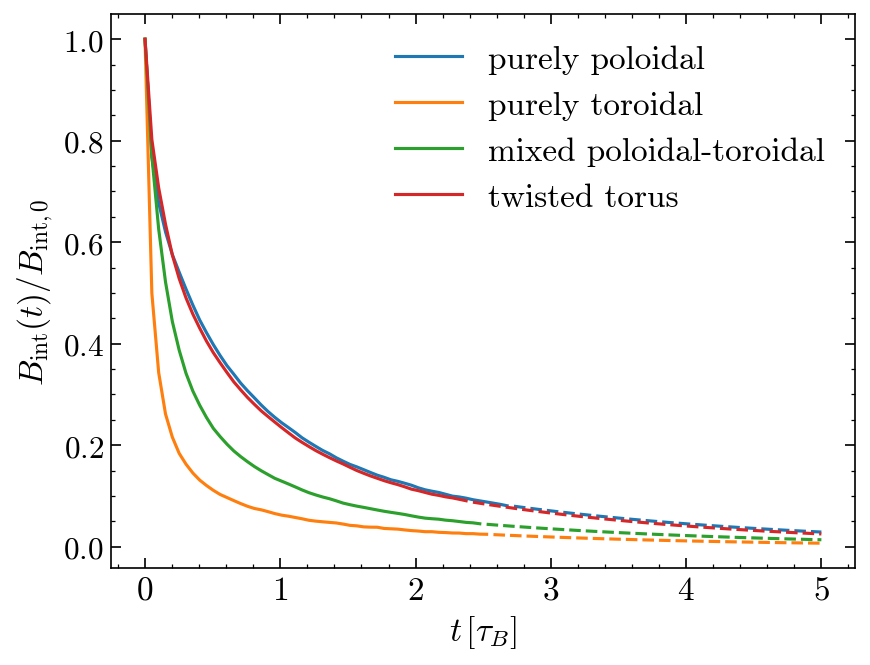}
\caption{Illustrative internal field-evolution models adopted in this work, based on the coupled core--crust magneto-thermal simulations of Ref.~\cite{Skiathas:2024jah}. The four curves correspond to different initial magnetic-field geometries: purely poloidal, purely toroidal, mixed poloidal-toroidal, and twisted-torus configurations. The solid segments show the simulation results, while the dashed segments indicate the late-time exponential extrapolation adopted beyond the simulation time span. The time axis is normalized by the characteristic evolution timescale $\tau_B$.
}
\label{fig:B_evolution}
\end{figure}

We relate their magnetic-energy evolution to the effective internal field by 
\begin{equation}
    B_{\rm int}(t)  = B_{\rm int,0}\left[\frac{E_B(t)}{E_B(0)}\right]^{1/2}.
\end{equation}
where $E_B(t)/E_B(0)$ is adopted from Fig.~6 in Ref.~\cite{Skiathas:2024jah}. In Fig.~\ref{fig:B_evolution} we show the evolution of $B_{\rm int}(t)$.
The simulations of Ref.~\cite{Skiathas:2024jah} are normalized to a fiducial field strength $B_0=10^{15}\,{\rm G}$ and $\tau_B\simeq2.2\,{\rm Myr}$.
Because the numerical tracks are available only up to a few $\tau_B$, we extrapolate the late-time evolution with an exponential decay matched to the final part of each track, as shown in Fig.~\ref{fig:B_evolution}. 
Although this extrapolation should not be interpreted as a reliable prediction at $t\gg\tau_B$, it has little impact on our estimates: systems with $t_{\rm m}\gg\tau_B$ have already lost most of the magnetic field and do not contribute appreciably to $P(f_{\rm br}>f_{\rm det,th})$.

This prescription should still be regarded as illustrative. The simulations of Ref.~\cite{Skiathas:2024jah} use time-independent microphysical parameters, corresponding in particular to a fixed temperature of $10^{9}\,\rm K$. A self-consistent thermal evolution, different initial field strengths, and possible effects associated with neutron superfluidity and proton superconductivity~\cite{Glampedakis:2010sk,Lander:2012zz,Graber2015,Passamonti:2017pme} could all modify the internal-field evolution and the effective retention time, while their quantitative impact on internal-field retention remains uncertain. 
Future studies should incorporate these effects and explore a broader range of initial conditions.

To evaluate $P(f_{\rm br}>f_{\rm det,th})$, we also need the merger delay-time distribution $p(t_{\rm m})$. 
As a reference delay-time model, we use the zero-kick model reported in Ref.~\cite{Stegmann:2025clo}. This model gives the median delay time, $t_{\rm m,median}=2.49\times10^{5}\,{\rm yr}$, and the 95th percentile, $t_{\rm m,95}=2.58\times10^{7}\,{\rm yr}$. 
Since Ref.~\cite{Stegmann:2025clo} does not provide an explicit distribution for $t_m$, we model $\log_{10} t_{\rm m}$ as a truncated normal distribution, with parameters chosen to reproduce the reported median and 95th percentile. 
This should be viewed as a reference parameterization rather than a unique prediction of the tertiary-driven channel.

\begin{figure*}%[h]--------------------------------------------------------------------------
\centering
\includegraphics[width=2.8in]{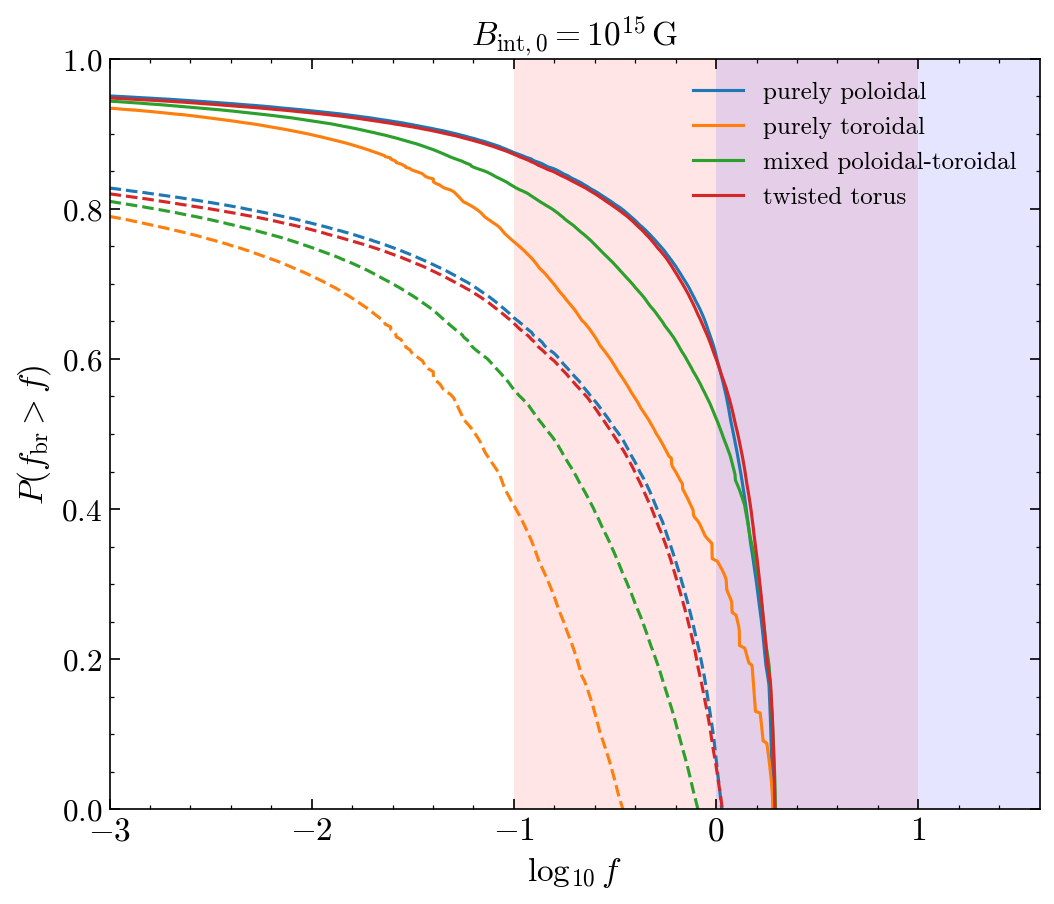}
\includegraphics[width=2.8in]{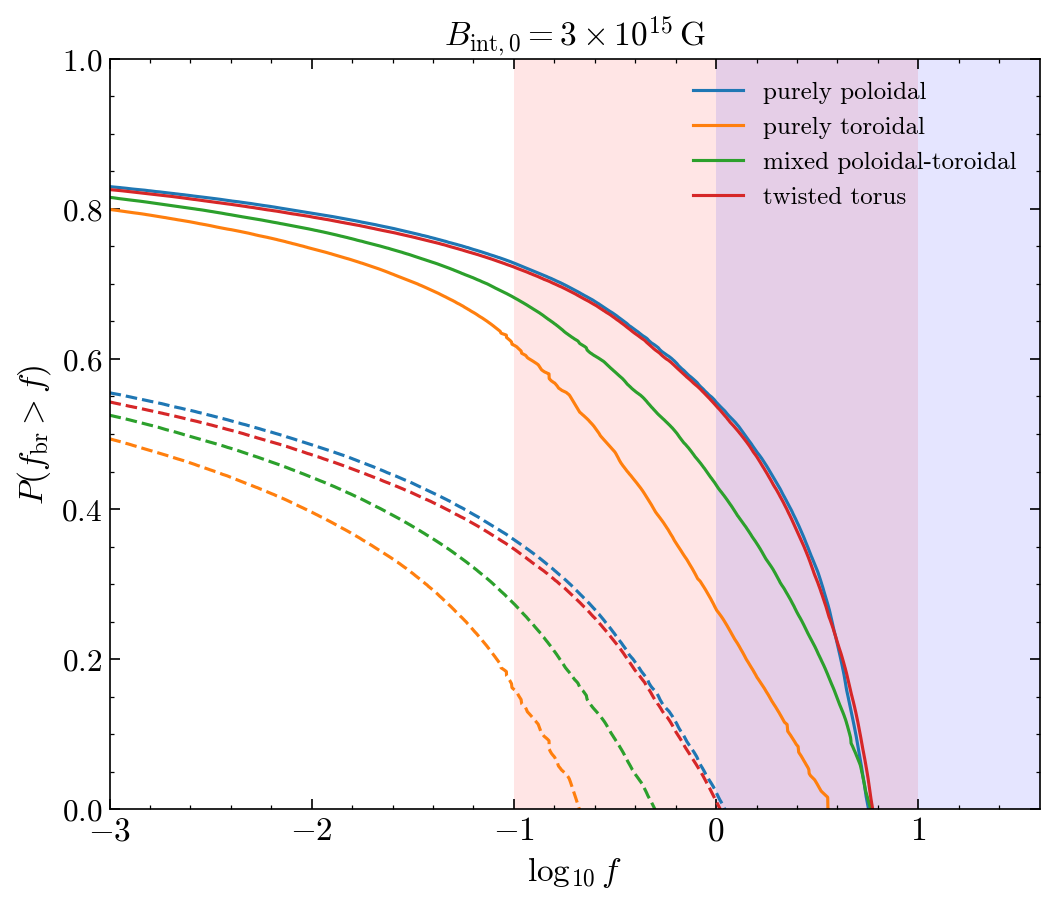}
\caption{Cumulative distribution of the breaking frequency, $P(f_{\rm br}>f)$.
The left and right panels correspond to initial internal magnetic fields $B_{\rm int,0}=10^{15}\,{\rm G}$ and $3\times10^{15}\,{\rm G}$, respectively.
Different colors denote different initial magnetic-field geometries adopted from Ref.~\cite{Skiathas:2024jah}: purely poloidal, purely toroidal, mixed poloidal-toroidal, and twisted-torus.
Solid curves show the benchmark choice with $\tau_B=2.2\,{\rm Myr}$ and $t_{\rm m,min}=10^{4}\,{\rm yr}$,
while dashed curves correspond to a more conservative choice, $\tau_B=0.22\,{\rm Myr}$ and $t_{\rm m,min}=4\times10^{4}\,{\rm yr}$.
The two shaded regions indicate the sensitive
frequency band corresponding to the DECIGO (0.1-10 Hz; red) and CE+ET ($\gtrsim$ 1 Hz; blue), respectively.
}
\label{fig:fbr}
\end{figure*}

We combine the merger delay-time distribution with the magnetic-field evolution models described above to generate the distribution of $B_{\rm int}(t_{\rm m})$ for a synthetic NSBH population.
The resulting internal field is then converted into the magnetic ellipticity and the corresponding breaking frequency using Eqs.~(\ref{eq:e-B}) and (\ref{eq:fbr_mag}).
Figure~\ref{fig:fbr} shows the cumulative probability $P(f_{\rm br}>f)$.

For our benchmark model, $B_{\rm int,0}=10^{15}\,{\rm G}$, $\tau_B=2.2\,{\rm Myr}$, and $t_{\rm m,min}=10^{4}\,{\rm yr}$,
we find that a fraction of order $\mathcal{O}(0.1)$ of systems retain sufficiently strong internal fields to reach $f_{\rm br}>1\,{\rm Hz}$, allowing the resonance-locking signal to enter the CE+ET band.
Combined with the choice of the reference $f_{\rm mag,0}=0.1$, this corresponds to $f_{\rm mag,merge}\sim\mathcal{O}(10^{-2})$.

Since this fraction is sensitive to both magnetic-field decay and the short-delay tail of the merger-time distribution, we also consider a shorter retention timescale $\tau_B=0.22\,{\rm Myr}$, and a larger minimum delay time $t_{\rm m,min}=4\times10^{4}\,{\rm yr}$, to assess the sensitivity of our results to these assumptions.
These cases are also shown in Fig.~\ref{fig:fbr}. We find that, even under these more conservative assumptions, the purely poloidal and twisted-torus configurations still retain a non-negligible probability of reaching $f_{\rm br}>1\,{\rm Hz}$ and entering the CE+ET band.

We also repeat the calculation for a larger initial field, $B_{\rm int,0}=3\times10^{15}\,{\rm G}$, as shown in Fig.~\ref{fig:fbr}.
Since $\tau_B$ characterizes mainly the core ambipolar
diffusion time scale~\cite{Skiathas:2024jah}, we rescale it as $\tau_B\propto B_{\rm int,0}^{-2}$ for this case. 
A larger $B_{\rm int,0}$ shifts the distribution towards higher breaking frequencies, with the upper end extending beyond $f_{\rm br}\sim5\,{\rm Hz}$.

We also note that, for both the benchmark and more conservative cases, a fraction of order $10^{-1}$ falls in the decihertz regime, $0.1$--$10\,{\rm Hz}$. This suggests that future decihertz detectors such as DECIGO~\cite{Kawamura:2011zz} may provide a more promising avenue for detecting these resonance-locking systems.

Several caveats should be taken into account. First, a fully predictive estimate of $f_{\rm mag,merge}$ would require integrating over the distribution of initial internal magnetic fields, whereas here we consider only a few representative values of $B_{\rm int,0}$ in order to assess the survival probability of strongly magnetized systems. Second, the conversion from $B_{\rm int}$ to ellipticity depends on the magnetic-field geometry through the coefficient $\kappa$ in Eq.~(\ref{eq:e-B}), which we fix to a fiducial value of $\kappa=1$. Different field geometries may therefore change the predicted $f_{\rm br}$ distribution. Together with the uncertainties in the magnetic-field evolution and merger-delay distribution discussed above, the results presented here should be regarded as illustrative estimates rather than robust population predictions.
A more comprehensive population study, including the initial internal-field distribution, magnetic geometry, and self-consistent field-evolution models, is needed to make more quantitative
predictions.

\end{document}